%% file: main.tex
\documentclass[10pt,twocolumn,letterpaper]{article}
\newcommand{\PAR}[1]{\vskip 4pt \noindent{\bf #1~}}
\usepackage{array}    % For better column alignment
\usepackage{multicol} % For spanning text across columns
\usepackage{comment}

\usepackage[]{cvpr}   

\definecolor{cvprblue}{rgb}{0.21,0.49,0.74}
\usepackage[pagebackref,breaklinks,colorlinks,allcolors=cvprblue]{hyperref}

\def\paperID{*****} % *** Enter the Paper ID here
\def\confName{CVPR}
\def\confYear{2025}

\title{Spot-On: A Mixed Reality Interface for Multi-Robot Cooperation}
\author{Tim Engelbracht, Petar Lukovic, Tjark Behrens, Kai Lascheit, Ren\'e Zurbr\"ugg,\\ Marc Pollefeys, Hermann Blum and Zuria Bauer\\
{\tt\small \{tengelbracht, plukovic, klascheit, zerene, pomarc, zbauer @ethz.ch\} and blumh@uni-bonn.de}}
\begin{document}
\maketitle
\vspace{-0.8cm}
\input{sec/0_abstract} 
\vspace{-0.8cm}
\input{sec/1_intro}
\vspace{-0.2cm}
\input{sec/2_related_work}
\vspace{-0.2cm}
\input{sec/3_method}
\vspace{-0.2cm}
\input{sec/4_user_study}

\vspace{-0.2cm}
\input{sec/5_conclusion}

\newpage

{
    \small
    \bibliographystyle{ieeenat_fullname}
    \bibliography{main}
}


\end{document}

%% file: sec/0_abstract.tex
\begin{abstract}
%Progress in mixed reality (MR) and robotics is enabling increasingly sophisticated forms of human–robot collaboration. Building on these developments, we introduce a novel MR framework that allows multiple quadruped robots to operate in semantically diverse environments via a HoloLens interface. Our system supports collaborative tasks involving drawers, swing doors, and higher-level infrastructure such as light switches. An underlying scene graph structure stores the evolving states of objects and their semantic connections as the robots interact with the environment. To assess both usability and design decisions, we conducted a user study with quantitative and qualitative feedback. Participants exhibited a steep learning curve, completing tasks notably faster after their first attempt; prior MR/VR familiarity further improved performance, whereas a brief textual tutorial had only modest impact. Qualitative responses highlighted preferences for default color schemes, an integrated transition button to the scene view, and an optional night mode, all guiding our further interface refinements. Overall, our approach provides an effective and intuitive framework for MR-based multi-robot collaboration in complex, real-world scenarios.
Recent progress in mixed reality (MR) and robotics is enabling increasingly sophisticated forms of human–robot collaboration. Building on these developments, we introduce a novel MR framework that allows multiple quadruped robots to operate in semantically diverse environments via a MR interface. Our system supports collaborative tasks involving drawers, swing doors, and higher-level infrastructure such as light switches. A comprehensive user study verifies both the design and usability of our app, with participants giving a "good" or "very good" rating in almost all cases. Overall, our approach provides an effective and intuitive framework for MR-based multi-robot collaboration in complex, real-world scenarios. A video demonstration of Spot-On can be found here: \href{https://photos.app.goo.gl/DvmatPNU96pDnMQN9}{video}.
\end{abstract}

%% file: sec/1_intro.tex
\section{Introduction}
\vspace{-0.1cm}
Recent advances in MR, VR and mobile robotics have opened the door to new possibilities in human–robot collaboration. Head-mounted displays, such as the HoloLens or Quest 3, have enabled intuitive remote operation of sophisticated robotic platforms, while concurrent developments in autonomous robot teaming have improved coordination among multiple agents. Existing work in the area - such as recent efforts on single-robot control using augmented reality headsets~\cite{holospot} or basic multi-robot teaming frameworks~\cite{chen} - demonstrates the potential of overlaying digital cues on the physical environment to enable more intuitive command and coordination. However, relatively few systems address rich multi-robot collaboration and object-specific interactions in cluttered or semantically diverse spaces. Moreover, most existing approaches only incorporate simplified digital representations of the environment, thereby limiting the robots’ ability to accurately localize, manipulate, and perform collaborative tasks that require an understanding of specific objects. Building on these insights, this paper presents a novel framework for controlling multiple quadruped robots in a shared workspace via a MR/VR interface. Semantic instance segmentation is applied to identify and distinguish different objects, while a scene graph encodes the relationships and affordances within the space. This digital clone of the environment allows operators to command robots to interact with various elements - ranging from drawers and swing doors to movable or more functional objects - with a high degree of autonomy. The semantic states of and connections between objects are updated within the scene graph while the robots interact with the environment.
\begin{figure}[t]
\vspace{-0.5cm}
    \centering
    \hspace{-0.1\columnwidth}
    \includegraphics[width=0.85\columnwidth]{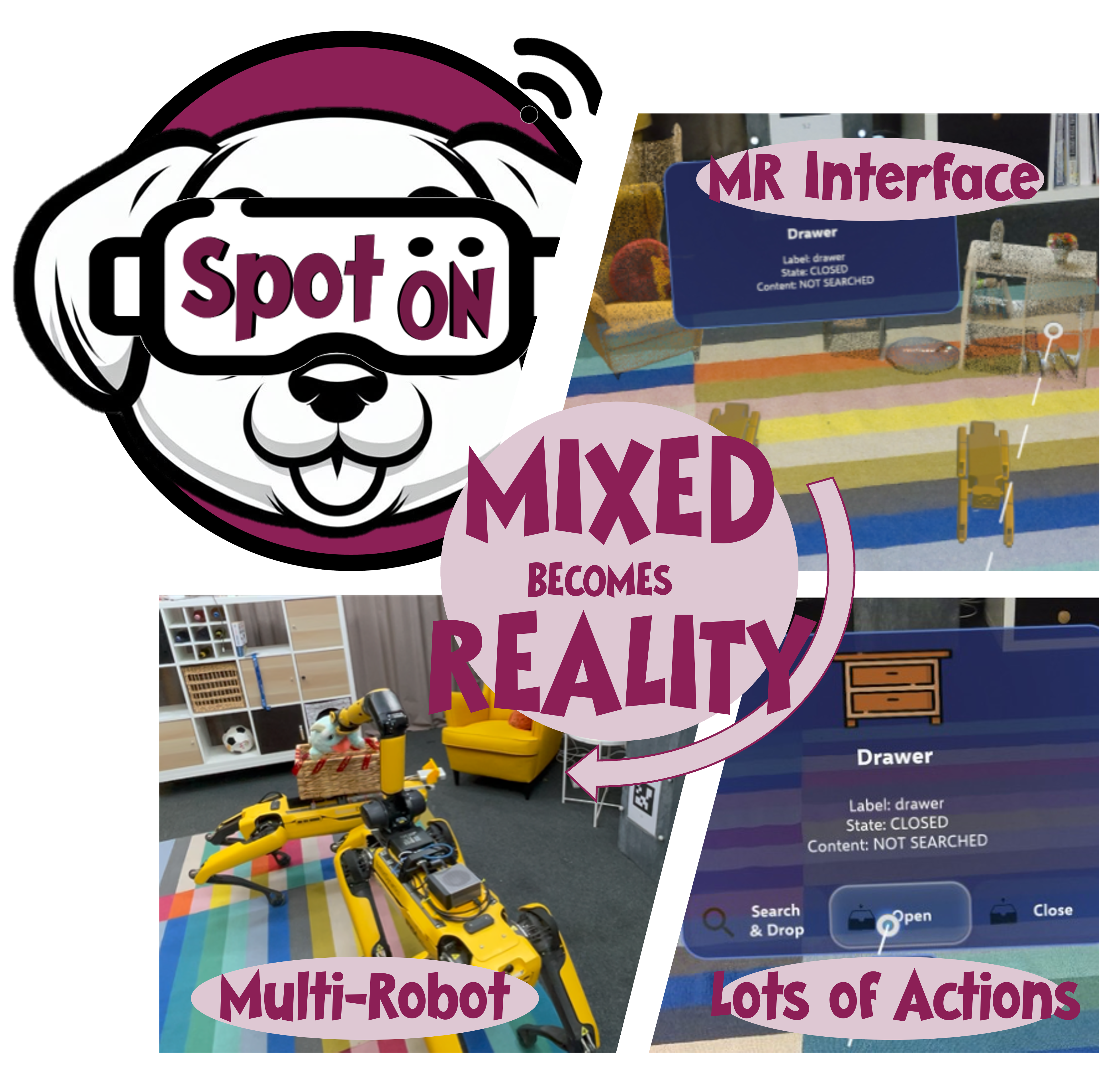}
    \caption{\textbf{Spot-On}. We introduce Spot-On, a Mixed-Reality application that allows Multi-Robot Control and Collaboration}
    \label{fig:teaser}
    \vspace{-0.7cm}
\end{figure}
Our approach extends the capabilities of previous systems~\cite{lemke2024} by enabling multi-robot collaboration on tasks such as object manipulation and joint inspection or activation of room infrastructure (e.g. light switches and lamps). In doing so, the proposed system demonstrates a more sophisticated paradigm for mixed reality–based control of multiple robots, one that is suited to complex and dynamic real-world scenarios. Contributions of this work include:
\begin{itemize}
    \item An interactive mixed reality interface for coordinated operation of multiple robots using a digital clone of the environment.
    \item A scene graph as the underlying environment model to enable robust object understanding and interaction.
    \item Mechanisms for advanced collaborative manipulation tasks (drawers, swing doors, dynamic drag-and-drop).
    \item Real-world demonstration of collaborative inspection tasks.
\end{itemize}

%% file: sec/2_related_work.tex
\section{Related Work}
\vspace{-0.1cm}
\begin{figure*}[h!]
    \centering
    \includegraphics[width=\textwidth]{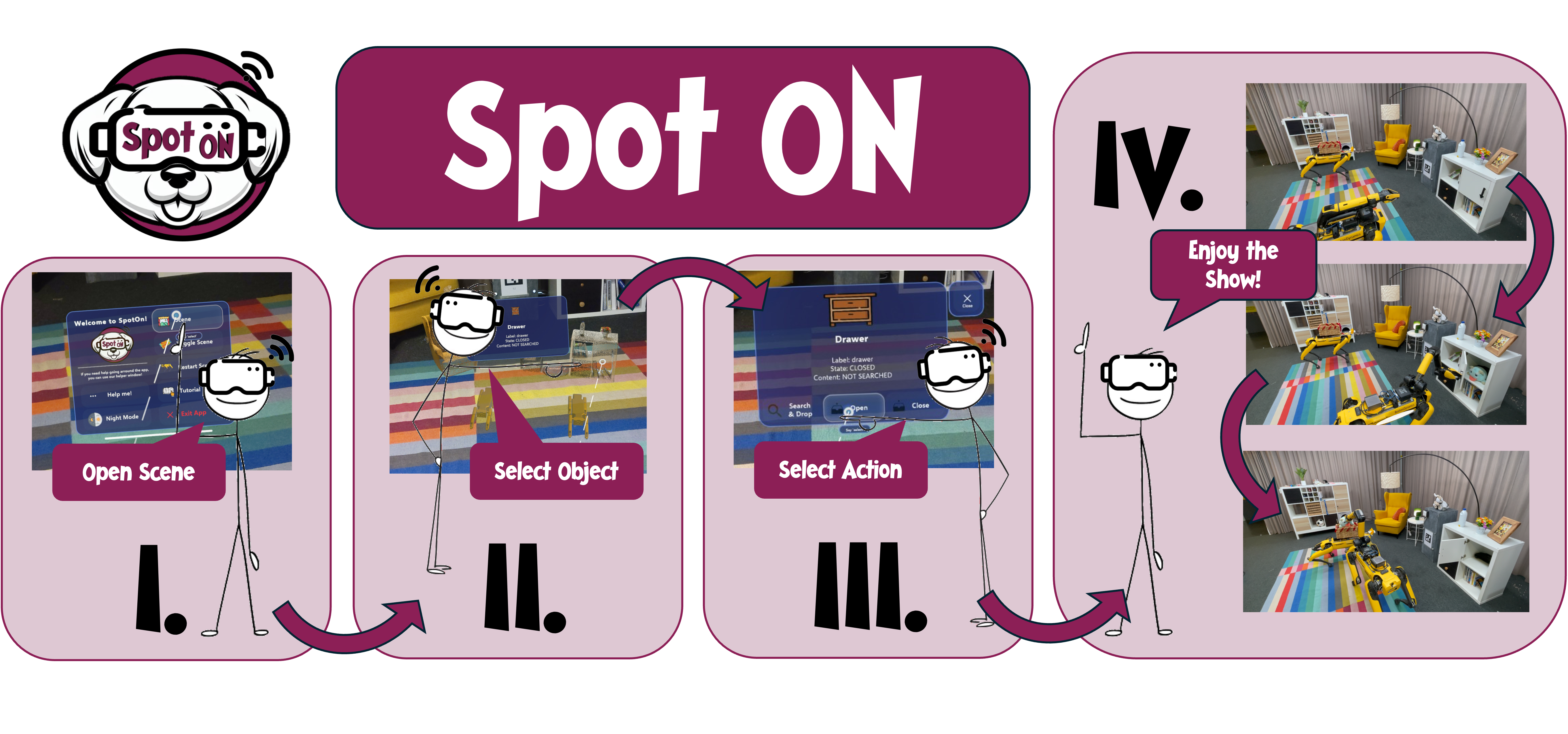} % Width set to \textwidth
    \vspace{-1.3cm}
    \caption{\textbf{Spot-On Overview.} Our user friendly top-level mixed-reality interface allows for intuitive robotic control. Starting from the main menu, the user opens the scene view (I.). The user is presented with an interactable 3D reconstruction of the scene. By hovering objects in the scene  and clicking them, objects can be selected (II.). Now, object specific actions are highlighted and can be selected at will (III.). Once the action is selected in the virtual user interface, the robots automatically start performing the task in the real scene (IV.).}
    \label{fig:double-column-image}
\end{figure*}

\PAR{Core tasks.}
We build our implementation on recent advancements in core robotics tasks such as scene reconstruction~\cite{lostfound, chen}, instance segmentation~\cite{mask3d, clip, openmask3d}, and object grasping~\cite{fang2023, ok-robot}. The Spot-Compose framework~\cite{lemke2024} integrates several of these techniques, providing a unified platform for robot-object interaction and manipulation. Specifically, it combines OpenMask3D~\cite{openmask3d} for open-vocabulary instance segmentation, AnyGrasp~\cite{fang2023} for arbitrary object grasping, and a custom method for robot repositioning to facilitate object handling. Additionally, Spot-Compose employs YOLOv8~\cite{Jocher_Ultralytics_YOLO_2023} for classifying drawers and handles and introduces a novel approach to estimate the axes of motion of drawers for automated opening. We extend this work by incorporating additional robotic capabilities and providing an intuitive and immersive mixed reality, and virtual reality, interface for remote robot control.

\PAR{Human-Robot interaction.}
Given the rapid development in mixed and virtual reality devices, Human-Robot Interaction (HRI) has naturally emerged as a novel research area. As early works~\cite{chen} suggest, 3D visualization in MR/VR enhance user experience and enable intuitive interaction between humans and robotic agents. Additionally,~\cite{iglesius2024mrnabmixedrealitybasedrobot} display HRI interface in an agent navigation setting using MRNaB. More recent work, HoloSpot~\cite{holospot} provides a compelling example of previously mentioned interface, implementing a digital twin of a scene in Unity and translating user inputs into robot commands via Spot-Compose. In doing so, HoloSpot facilitates object drag-and-drop functionality on the Microsoft HoloLens~\cite{hololens}, allowing users to specify actions in mixed reality, which are then executed in the real world by the robot. While HoloSpot demonstrates robust capabilities in object picking and placing, it has two primary limitations. First, it does not account for relationships between objects within the scene, which are often critical for contextual manipulation tasks, such as operating light switches. Second, it supports only a single robot, limiting its applicability in collaborative scenarios. To address these limitations, we propose a novel mixed reality (and virtual reality) interface that incorporates elements of SpotLight~\cite{engelbracht2024}, enabling robots to assess and modify the states of lamps and light switches (i.e. on or off) based on user commands. %Furthermore, we develop a novel method for coordinating two robots within a shared scene to collaboratively execute pick-and-place operations and more complex object manipulation tasks.

\PAR{Scene Graphs.} 
Scene understanding is at the core of our work. Similarly to SpotLight~\cite{engelbracht2024}, we utilize scene graphs~\cite{conceptgraphs} in order to facilitate interactive perception tasks. This is the crucial part of our work since our goal is to allow the agent to interact with the environment and acquire knowledge through this interaction~\cite{DBLP:journals/corr/BohgHSBKSS16}. Previous works on scene graphs and interactive perception were mostly utilized to learn object articulations~\cite{nie2023structureactionlearninginteractions, hsu2023dittohousebuildingarticulation, jiang2022dittobuildingdigitaltwins, 10.1109/ICRA46639.2022.9812430, 7139655, inproceedings}, in our work we employ scene graph as a medium to store scene information and object relationships of a dynamically changing environment, as well as to delegate tasks in the context of human-robot interaction. During interaction, changes in the environment are monitored and tracked by capturing them in a dynamic scene graph. Therefore, we can learn new relations between objects and carry out dynamical updates on the scene graph, such as in~\cite{rosinol20203ddynamicscenegraphs, wang2024articulated}.

%% file: sec/3_method.tex
\section{Method}
\vspace{-0.1cm}
\label{method}
Our method relies on using a scene graph as a central data structure. This approach offers several advantages: Firstly, the scene graph provides a 3D representation that the user can easily interact with using an MR or VR interface, such as pointing and clicking. This method is more reliable than open language interfacing, which is prone to misinterpretation. Secondly, the scene graph accommodates dynamic scene changes, such as lamps turning on and off, capturing these changes in real-time and storing them for later use. Lastly, it simplifies data storage and management, enabling easy information sharing between robots and human operator. For instance, if one robot changes something in the environment, the other robots are immediately aware of this change through the scene graph, effectively serving as a shared brain. Based on these benefits, we propose building the scene graph offline and then integrating it into an MR application to allow a human to interact with the scene remotely. This setup enables the delegation of tasks to robots. As the robots interact with the scene, they dynamically update the underlying representation, which is visually relayed to the human user through the MR/VR interface. In the following sections, we will describe the scene graph structure, the MR/VR application, and the robotic tasks in detail.

\subsection{Scene Graph}
\label{subsec:scene-graphs}
\vspace{-0.1cm}
While open-language instance segmentation~\cite{openmask3d} offers higher flexibility, we identified the resulting data structure (point cloud with open language features) as less suitable for efficient and robust robotic downstream tasks. 
Compared to previous approaches~\cite{holospot, chen}, we rely on an object-centric scene graph~\cite{lostfound} that captures semantic relations between those instances, which inherently offers semantic scene understanding. Hence, we can effectively employ this scene graph for querying the scene via the the MR interface. We introduce the scene graph $G = (V, E)$ as a set of vertices $V=\{v_{1},...,v_{n} | v_i = (\xi_i, c_i, \mathcal{P}, \Phi_i, s_i)\}$ representing the scene's objects and a set of edges, representing the object's spatial relationships. Here, each vertex is defined by a pose $\xi_{i}=(\vec o_{c}, \vec n_{c})$, a sematic class $c_i \in \{1,...,N_c\}$, a point cloud $ \mathcal{P}_i = \{ p_n \mid p_n = (x_n, y_n, z_n, a_n), \, n = 1, \ldots, N \}$, a set of motion primitives $\Phi_i =\{\phi_{j}\}_{j=1}^{K}$ and an optional state $ s_i \in \{\texttt{on}, \texttt{off}, \texttt{none}\}$. Note that we define motion primitives for functional elements in accordance with ~\cite{scenefun3d}. To construct the initial scene graph, the scene is recorded using an iPad lidar scanner, where a point cloud as well as a sequence of posed RGBD images is captured. We obtain the nodes by segmenting the scene using the semantic segmentation method Mask3D ~\cite{mask3d}. Since this method fails to detect smaller, functional elements such as light switches, we detect them in the posed RGBD sequence using a fine-tuned YOLOv8 ~\cite{Jocher_Ultralytics_YOLO_2023}, lift into 3D and associate instances by clustering. We establish spatial relationships based on nearest neighboring. Interactions of a robot as explained in \ref{subsec:tasks} allow us to associate light switches with lamps, as well as to check the states of these lamps and drawers. To enable efficient transfer between our central compute unit and the Mixed Reality device, we save the scene graph as a lightweight JSON object.
\vspace{-0.1cm}
\subsection{Communications}
\begin{figure}[h!]
    \centering
   \includegraphics[width=0.9\columnwidth]{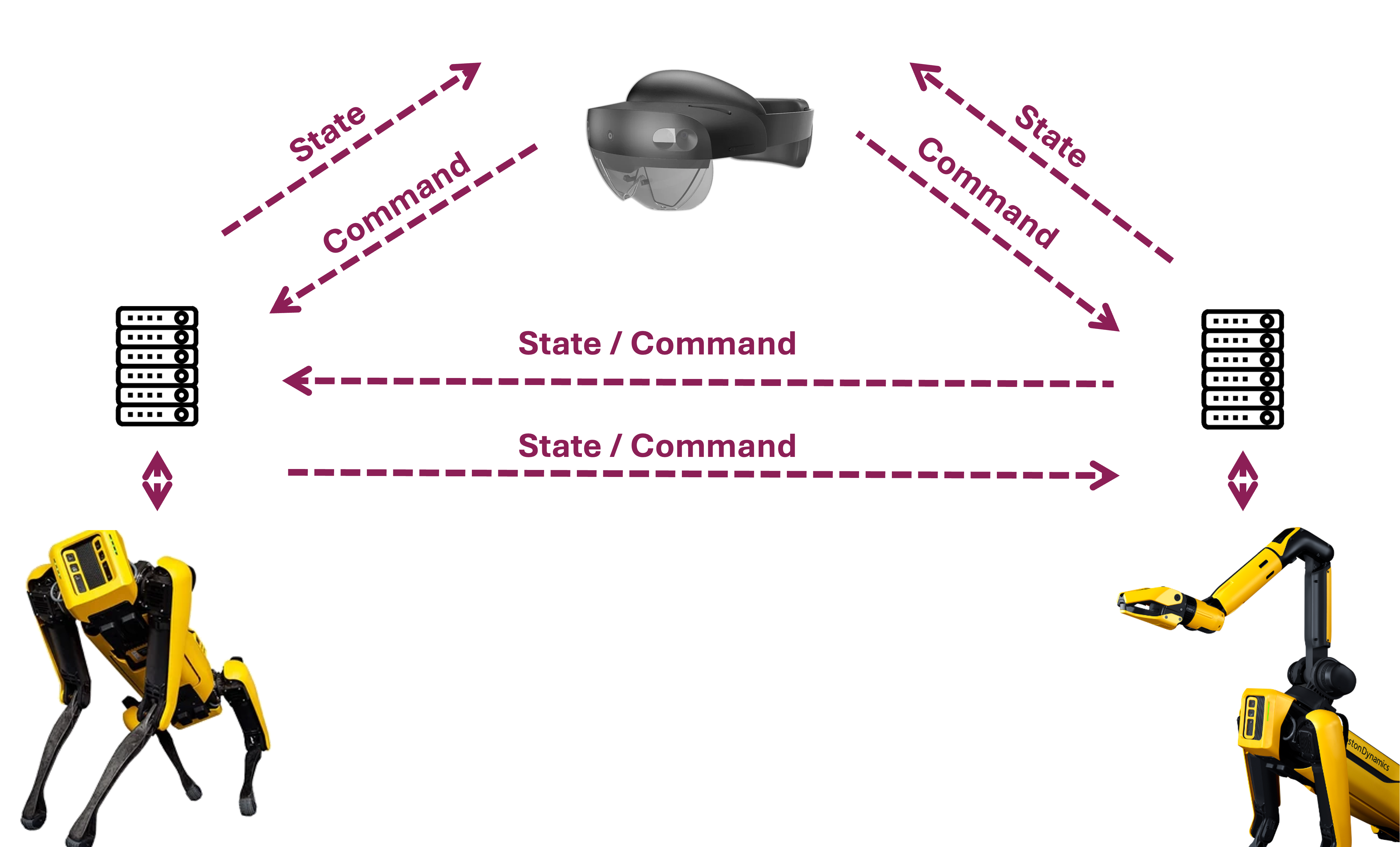} % Replace "example-image" with your image file nam
    \caption{\textbf{Networking.} The networking infrastructure relies on one server per robot. The HoloLens posts commands for the robots to execute, while the robots post their states and changes they have made in the environment. We query these changes, so that the scene graph representation in the user interface stays up-to-date.}
    \label{fig:networking}
    \vspace{-0.3cm}
\end{figure}

%lets talk about server infrastructure both on the robot side and on the Hololens side
\vspace{-0.1cm}
As shown in \cref{fig:networking}, our wireless networking infrastructure relies on a server for each robot. Through these servers, the robots both post their own state $x_{robot}=(x_{battery}, \ p, \ \theta, \ x_{status})$ as well as changes to the state of an object $y_{object}=(y_{ID}, \ y_{state})$ in the scene they caused by interactions. We denote $x_{battery} \in [0,1]$ as the battery status, $p \in \mathbb{R}^2$ as the 2D coordinates in the map, $\theta \in [0, \ 2\pi]$ as the orientation and $x_{status} \in \{ \texttt{IDLE, BUSY}\}$ as the busy status. Furthermore, $y_{ID} \in \mathbb{N}_0$ corresponds to the object ID in the scene graph and $y_{state} \{0, \ 1\}$ to the binary state of the object. Binary states can be represented as $(0, \ 1)$, which differ in interpretations depending on the context:
\begin{itemize}
    \item For a drawer: $(0, \ 1)$ corresponds to \texttt{(open, closed)}.
    \item For a lamp: $(0, \ 1)$ corresponds to \texttt{(off, on)}.
\end{itemize}
On the MR/VR devices' side, each scene graph object queries both servers at $10$ Hz for state changes of its physical counterpart. Once a change is posted by one of the robots, the scene graph on the MR/VR device and thereby the scene in the user interface gets updated.
In addition to state information, a job queue is hosted on the servers, routing user commands to the corresponding robot. To execute Multi-robot tasks (See \ref{subsec:tasks}), the robots can also send commands to each other and query their respective states. 
%todo: notation for jobs/commands?

\vspace{-0.1cm}
\subsection{Mixed Reality App}
\begin{figure}[h]
    \centering
    \begin{subfigure}[b]{0.49\columnwidth}
        \centering
        \includegraphics[width=\linewidth]{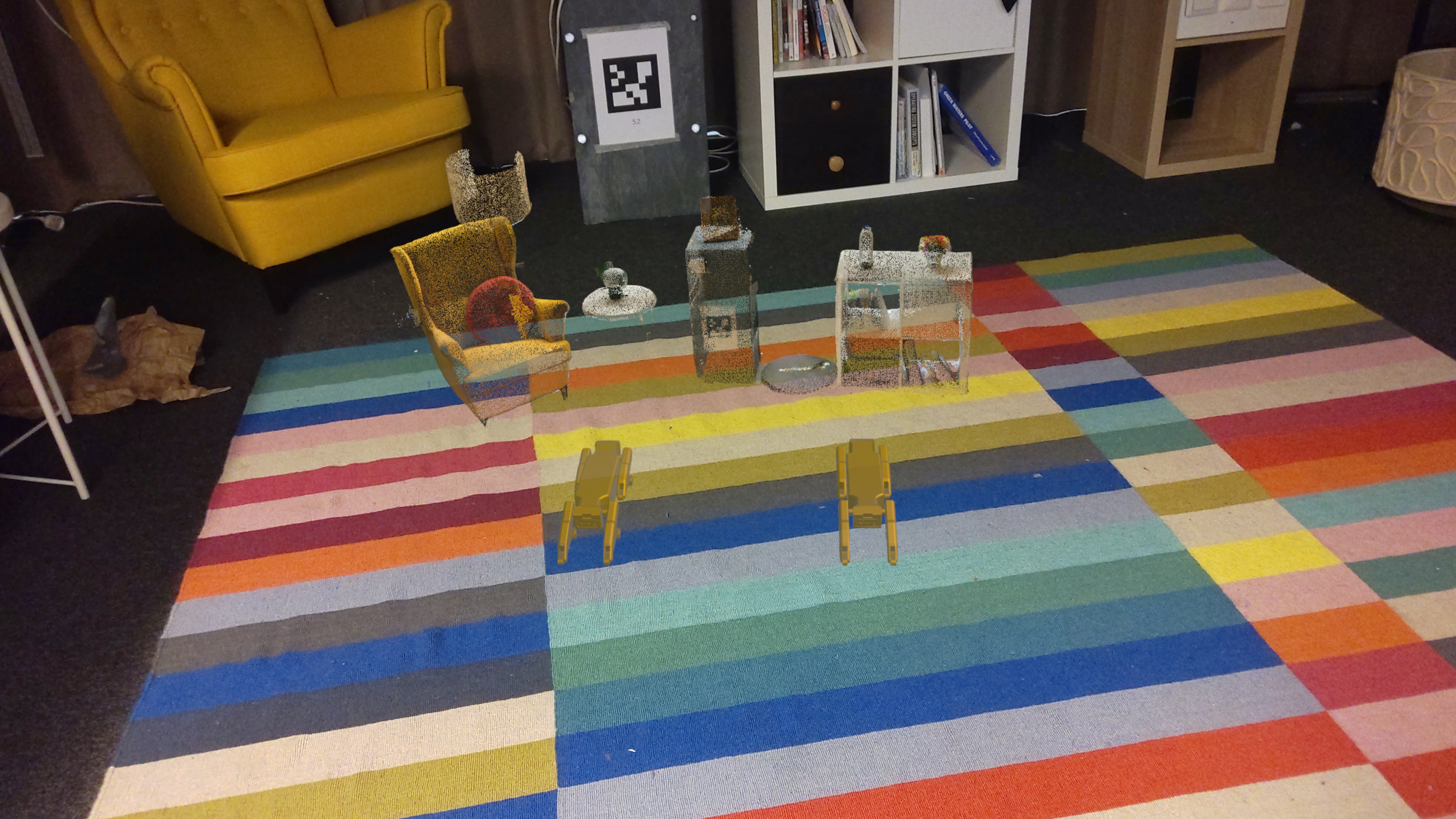}
    \end{subfigure}
    \hfill
    \begin{subfigure}[b]{0.49\columnwidth}
        \centering
        \includegraphics[width=\linewidth]{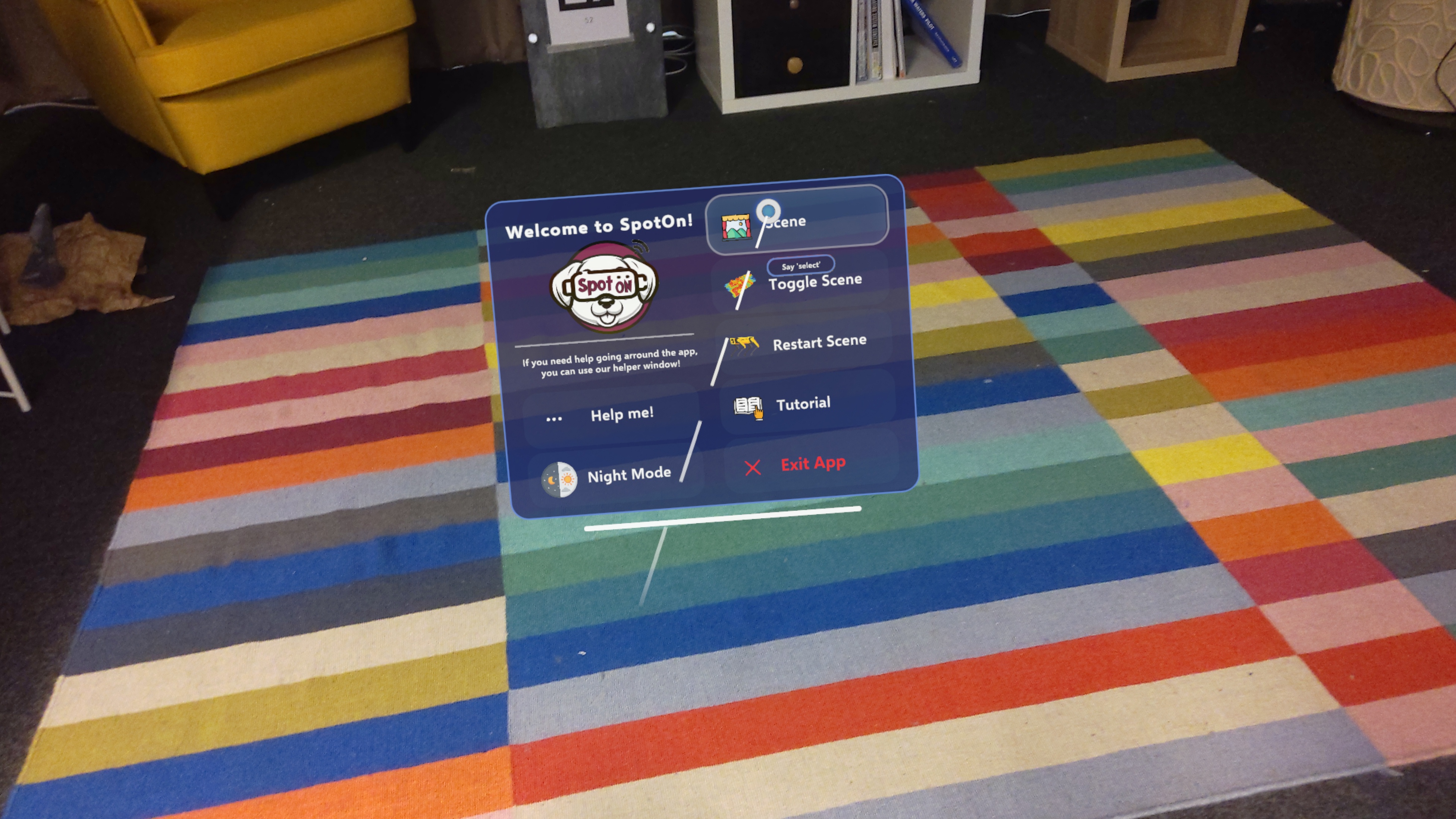}
    \end{subfigure}
    \caption{\textbf{Scene and Menu.} Figure shows scene (left) and start menu (right) as they are seen by user in the MR/VR device.} 
    \label{fig:scene-menu}
    \vspace{-0.5cm}
\end{figure}

% petar not sure what goes here exactly just give a short paragraph on the features and stuff
%In this section we will guide you through functionalities of our app and give you insight into full process of interacting with the real scene.
\vspace{-0.1cm}
We introduce a mixed reality app to interact with both the environment and robots. Built in Unity, it can be deployed on both pure Mixed Reality devices like Microsoft HoloLens, but also Virtual Reality devices like the Meta Quest.
Our app consists of two main parts, as shown in \cref{fig:scene-menu}. First, a home menu housing the app's core functionalities, and second, a $3D$ scene for interacting with different objects. Upon start-up, the app will display a splash screen displaying the Spot-On logo, followed by a widget asking the user whether they would like to use our tutorial or not. After accepting, the user will be shown the collapsible helper window, followed by the main menu. If the user declines the tutorial, they are directed towards the main menu immediately. At this point, the voice command feature is also available. In the following, we will give in-depth explanations on the tutorial, home menu and scene view. Additionally, we present scene interactions and the voice command feature.

\begin{comment}
\begin{table*}[t]
    \centering
    \begin{tabular}{>{\raggedright\arraybackslash}p{0.25\textwidth} p{0.25\textwidth} p{0.4\textwidth}}
        \toprule
        \textbf{Home menu command}& \textbf{Voice command} & \textbf{Description} \\ \midrule
        Scene button              & \textit{"Show scene"} & Opens $3D$ scene \\
        Toggle scene button       & \textit{"Toggle scene"} & Toggles scene colors \\ 
        Tutorial button           & \textit{"Open tutorial"} & Opens tutorial window \\ 
        Help me button            & \textit{"Help me"} & Opens helper window \\ 
        Restart scene button      & \textit{"Restart scene"} & Restarts scene to original orientation and distance \\
        Day/Night mode button     & \textit{"Day/Night mode"} & Switches between day and night mode \\ 
                                  & \textit{"Close"} & Closes current prompt \\ 
                                  & \textit{"Open menu"} & Opens main menu \\ \bottomrule
    \end{tabular}
    \caption{\textbf{Commands.} Table displays list of available commands in a main menu and their corresponding voice commands. Considering that voice command list is longer, commands that do not have their corresponding menu button are left blank.}
    \label{tab:menu-voice-commands}
    \vspace{-0.5cm}
\end{table*}
\end{comment}

\vspace{-0.1cm}
\PAR{Tutorial} We implement a short tutorial to help inexperienced users familiarize themselves with the app and its functionalities. For the inital release of the app, the tutorial consists of textual explanation for interacting with the MR/VR interface and the exact functionalities of the app. The tutorial is arranged as a circular buffer of text prompts containing information about the following: general interaction, start menu, interactable scene elements, and voice commands. Here, all the functionalities are explained in detail. The user can activate the tutorial anytime later both from start menu and by using voice commands (See \cref{fig:scene-menu}).

% Technically now we have a video as a tutorial and the old textual explanation is given inside of a helper window

\begin{figure}[h!]
    \centering
    \setlength{\tabcolsep}{1pt} % Reduce space between columns
    \renewcommand{\arraystretch}{0.8} % Reduce space between rows
    \begin{tabular}{@{}ccc@{}} % Remove padding around the table
        \includegraphics[width=0.33\columnwidth]{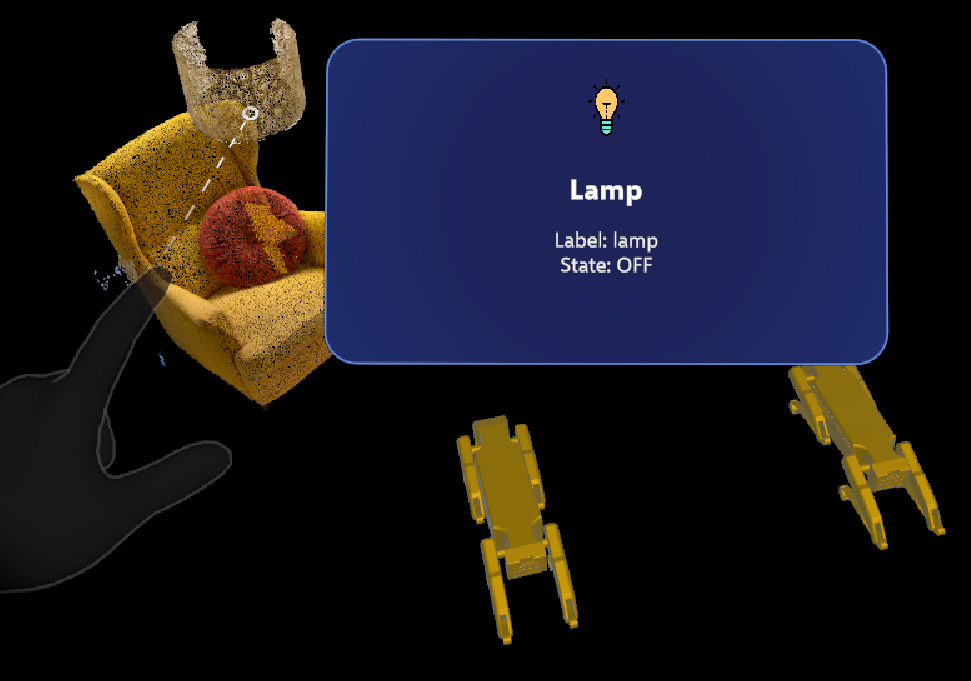} &
        \includegraphics[width=0.33\columnwidth]{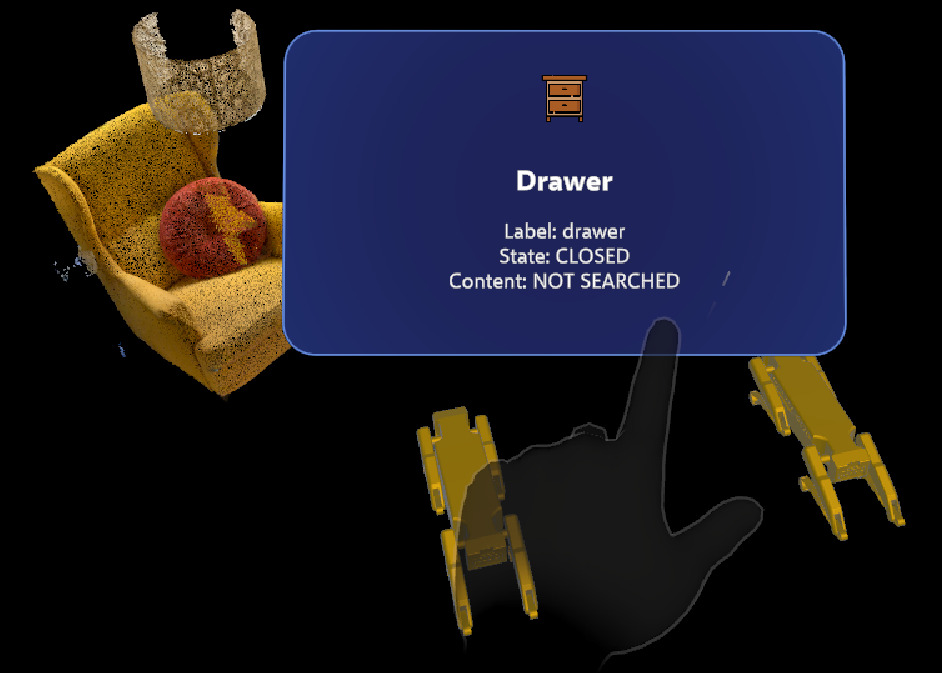} &
        \includegraphics[width=0.33\columnwidth]{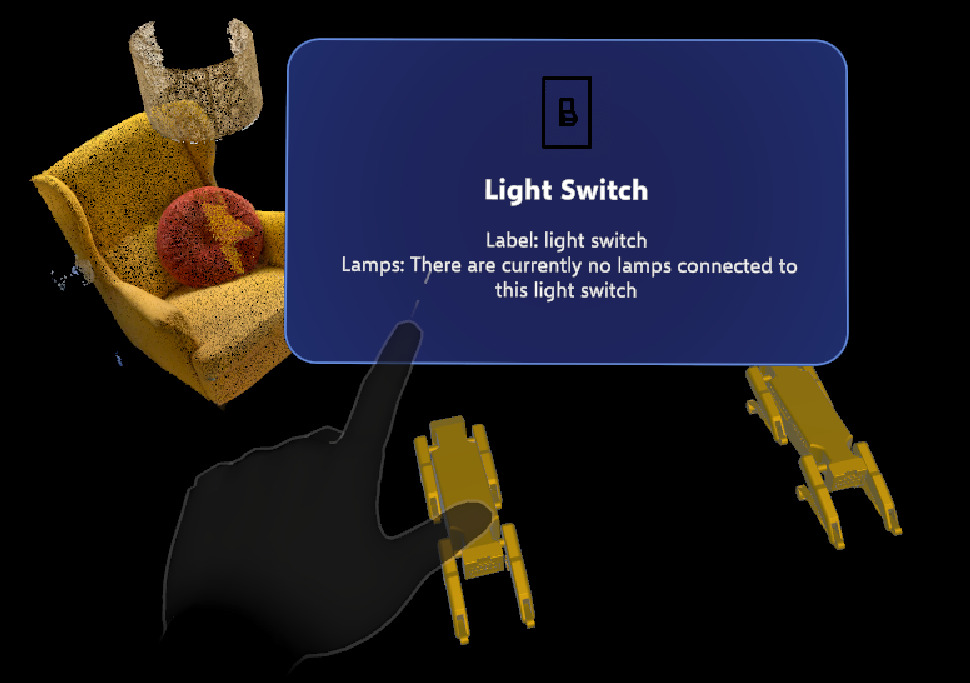} \\
        \includegraphics[width=0.33\columnwidth]{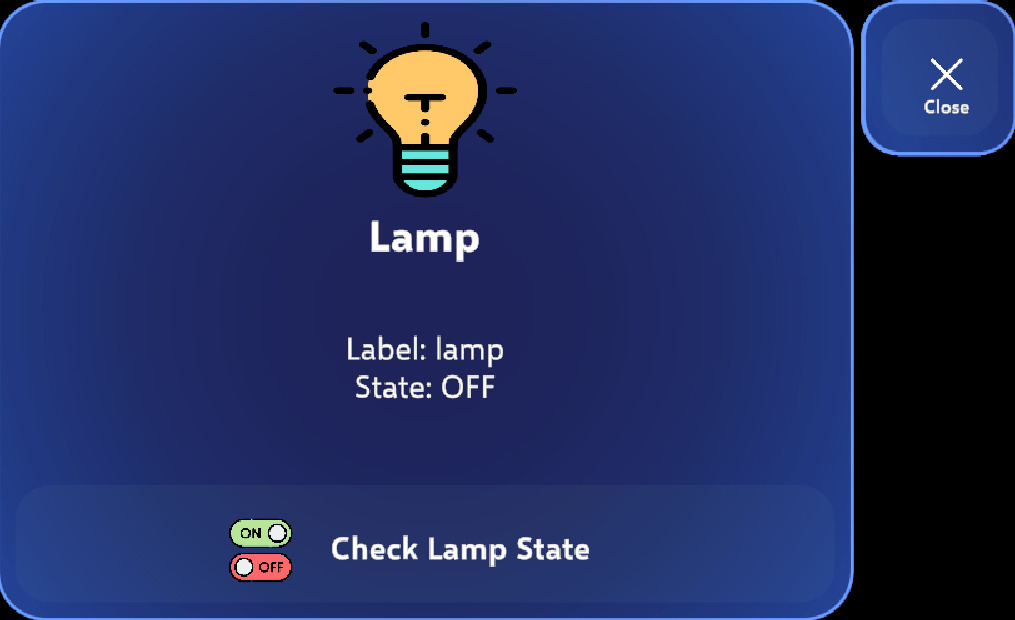} &
        \includegraphics[width=0.33\columnwidth]{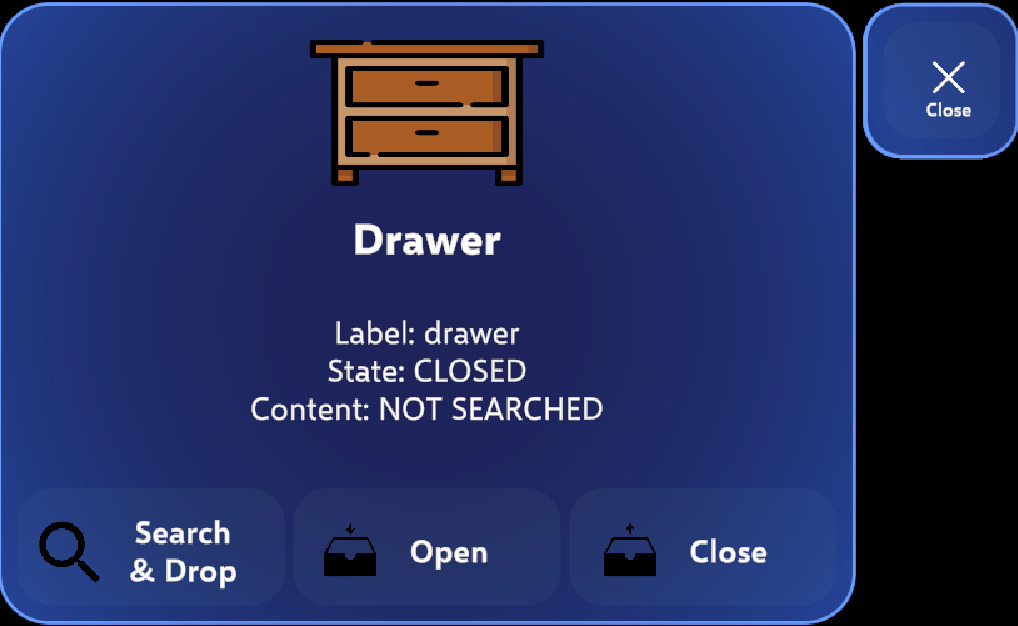} &
        \includegraphics[width=0.33\columnwidth]{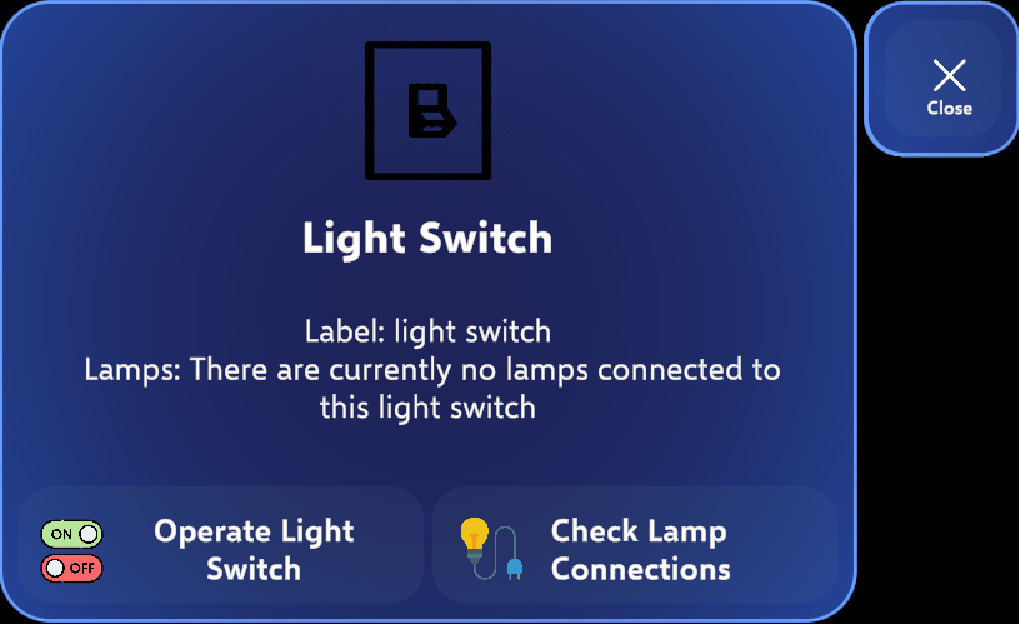} \\
    \end{tabular}
    \caption{\textbf{Interface.} Figure depicts the app's response to user interaction by hovering and clicking on different object types. The first row shows hovering events (popups), and the second row shows clicking events (prompts). From left to right, interactions are with a lamp, drawer, and light switch.}
    \label{fig:popup-prompt}
    \vspace{-0.5cm} % Reduce space below the figure
\end{figure}

\vspace{-0.1cm}
\PAR{Home Menu} Our main menu is shown in \cref{fig:scene-menu} (right) and \cref{fig:day-night-mode}. The home menu let the user access functionalities such as scene view or tutorial. A comprehensive list of functionalities is given in \cref{tab:menu-voice-commands}. All mentioned functionalities are also available through voice interface which will be explained later.
\vspace{-0.1cm}
\PAR{Scene View} In the scene view, as in \cref{fig:scene-menu} (left), the user can interact with the scene, with both objects and robots. The user enters scene view by either clicking on the 'Scene' button in the main menu or by using the voice interface. The Scene view displays a 3D point cloud representation of a real, scanned environment (\cref{subsec:scene-graphs}). Like all app widgets, the scene view follows the user's head movements. However, it is less sensitive compared to prompts to minimize users fatigue, particularly in the eyes and neck.
\vspace{-0.1cm}
\PAR{Interactions} Once in the scene view, the user can interact with all robots and objects in the scene. We discern two basic interaction types: hovering and selecting. Whenever the user hovers over an interactable object, a temporary prompt will appear and show basic information about the currnet object. As soon as the user's pointer is outside of the object bounding-box, this temporary pop-up disappears. To select a hovered object, the user performs a clicking action. This will open a custom object widget which allows the user to execute the functions explained in section \ref{subsec:tasks}. A comprehensive set of custom object widgets is displayed in \cref{fig:popup-prompt}. The user may leave the scene view and enter the main menu by selecting any point that is not an interactable object. Note that we deliberately choose to handle all interactions through clicking rather than natural language. While scene graph representations can indeed be queried using natural language when paired with a large language model (LLM) ~\cite{conceptgraphs}, we opt for clicking due to its lower ambiguity.
\vspace{-0.1cm}

\PAR{Voice Interface}
\label{voice-interface}
An additional useful way of interacting with the app are the voice commands. We implement voice commands as a result of user feedback during our pre-study (see \cref{subsec:pre-study}). Voice commands reduce the number of clicking interactions a user has to perform and thereby leading to a smoother user experience. Note that voice commands are used only to navigate the app, not to issue commands to the robots. The reason for this approach is to prevent accidental voice commands that could lead to unintentional physical action. Additionally, we do not allow critical voice commands such as exiting the app. Furthermore, while voice commands are permitted for certain important but less critical actions, such as restarting the scene, these actions require the user to confirm by clicking an additional prompt to prevent unintended operations. A comprehensive list of all voice commands is provided in \cref{tab:menu-voice-commands}.
\vspace{-0.1cm}
\PAR{Day/Night Mode}
As an additional functionality, we enable the switching between lower contrast day mode and a high contrast night mode. Modes are automatically activated depending on the time of the day, however, user can change mode using both start menu and voice commands.

\begin{table}[h]
    \centering
    \begin{tabular}{>{\raggedright\arraybackslash}p{0.4\columnwidth} p{0.5\columnwidth}}
        \toprule
        \quad \textbf{Voice command} & \textbf{Description} \\ \midrule
        \checkmark \ \textit{"Show scene"} & Opens $3D$ scene \\
        \checkmark \ \textit{"Toggle scene"} & Toggles scene colors \\ 
        \checkmark \ \textit{"Open tutorial"} & Opens tutorial window \\ 
        \checkmark \ \textit{"Help me"} & Opens helper window \\ 
        \checkmark \ \textit{"Restart scene"} & Restarts scene to start pose \\
        \checkmark \ \textit{"Day/Night mode"} &  Switches to day/night mode \\ 
        \quad \textit{"Close"} & Closes current prompt \\ 
        \quad \textit{"Open menu"} & Opens main menu \\ \bottomrule
    \end{tabular}
    \caption{\textbf{Commands.} Table displays list of available commands in a main menu and their corresponding voice commands. Considering that voice command list is longer, commands that have their corresponding menu button are indicated with a checkmark.}
    \label{tab:menu-voice-commands}
    \vspace{-0.3cm}
\end{table}

\vspace{-0.2cm}
\subsection{Robotic Collaboration for Multi-Robot Tasks} 
\label{subsec:tasks}
\vspace{-0.1cm}
\PAR{Agents} Our method is deployed on two Boston Dynamics Spot robots. The first robot is equipped with an arm and gripper, whereas the other robot does not have grasping capabilities. In the following, we will refer to the robot with arm simply as "Fluffy" and the robot without arm as "Softy". Due to asymmetric capabilities, all tasks that require gripping actions are automatically delegated to Fluffy. To increase Softy's utility, we equip it with an RGB camera and a basket fixed to its back. This allows the robot to observe its environment and carry objects.
\vspace{-0.1cm}

\PAR{Tasks} For collaboration we define a set of collaborative tasks that can be decomposed into a single robot tasks. Before elaborating on multi-robot tasks, we will first discuss single robot tasks. The user commands the robot to interact with a specific object through interactive buttons that appear upon hovering and clicking an interactable object in the mixed reality interface. After interaction, all state changes are posted on the server and the scene graph is updated accordingly. The tasks are listed as follows:
\begin{itemize}
    \item \textbf{Drawer Interaction}To compute the mption primitive for opening and closing of linear drawers, we utilize the Spot-Compose ~\cite{lemke2024} framework. The motion origin is detected as the handle using a fine-tuned YOLOv8 ~\cite{Jocher_Ultralytics_YOLO_2023}, while the axis of motion is computed as normal of the drawer plane.
    \item \textbf{Swing Door Interaction} Similar to the drawer detection, the handle is detected using a YOLO model. If the handle is detected to be eccentrically placed on the door, we compute a rotational 3D trajectory, similar to ~\cite{engelbracht2024}. 
    \item \textbf{Light Switch Operation} We have the agent operate light switches using the SpotLight ~\cite{engelbracht2024} framework. Here, the robot first refines the 3D position of the light switch and thereafter predicts the affordance, i.e. the type of interaction necessary for operation. Lastly, it estimates a 3D motion needed for operation. 
    \item \textbf{Grasp Object} As done in ~\cite{lemke2024}, object grasps are calculated using AnyGrasp ~\cite{fang2023}, which computes possible grasps based on the 3D segmentation mask of the object.
    \item \textbf{State Check} Depending on the object type, one can assign states to the objects in the scene. To this end, we define a set of states for drawers $x_{drawer}=\{\texttt{open}, \texttt{closed}\}$ and lamps $x_{lamp}=\{\texttt{on}, \texttt{off}\}$. Taking the idea from ~\cite{engelbracht2024}, the agents capture images of the objects and utilize the GPT4 API ~\cite{OpenAI2025} to infer the current state of the captured object.
\end{itemize}
Having defined single robot tasks, we will define multi-robot tasks as a coordinated sequence of sub-tasks as described in the single robot tasks. Combining a number of subtasks into a single multi-robot command leads to better usability since only one button has to be pushed in the interface. Similarly to single robot tasks, these tasks are sent to the agents via a button bound to a specific object. The multi-robot tasks are listed below.
\begin{figure}[h!]
    \centering
    \includegraphics[width=0.9\columnwidth]{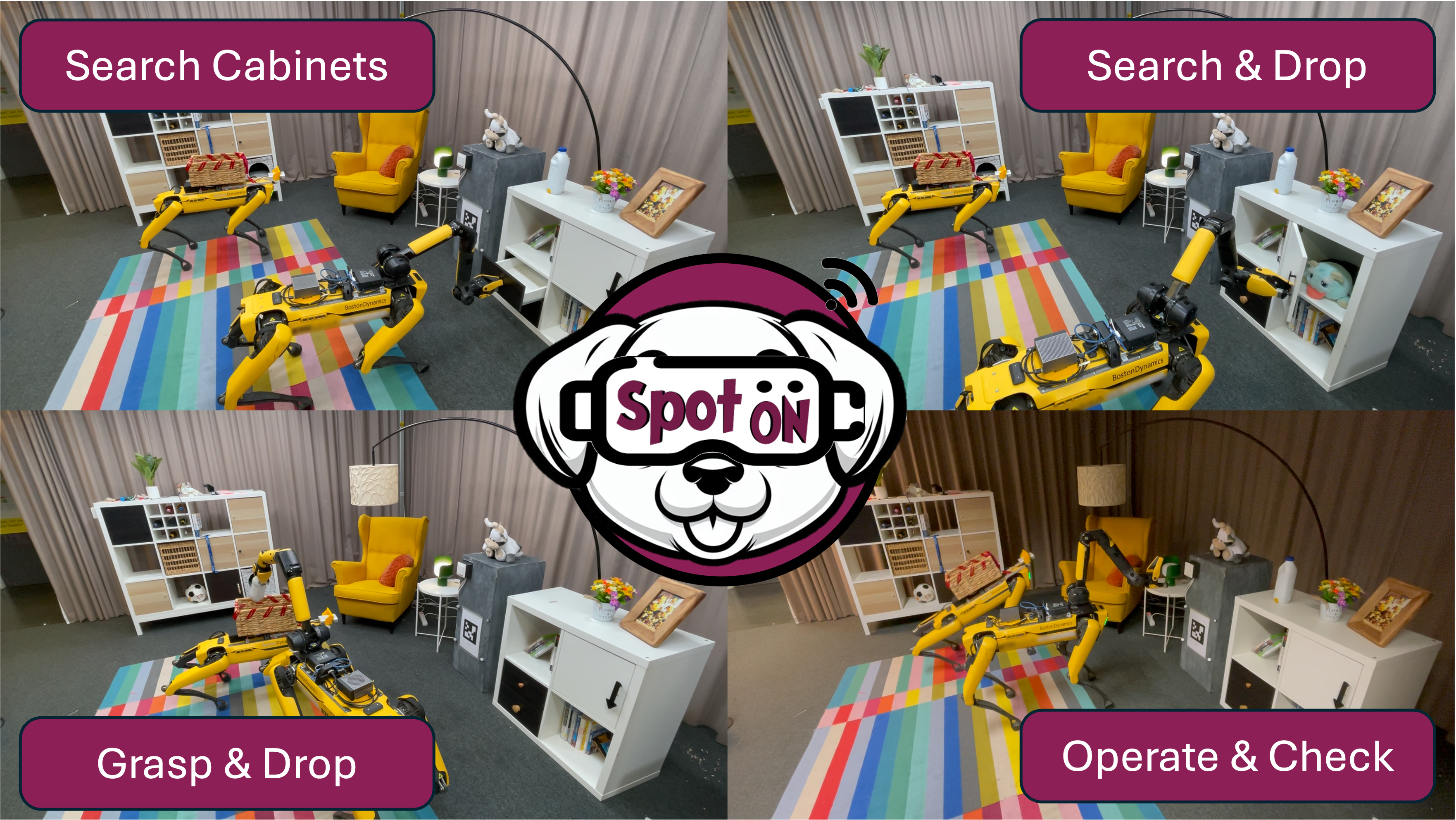} % Replace "example-image" with your image file nam
    \caption{\textbf{Robots in action.} Spot-On offers a wide range of robotic interactions: Opening drawers (top left) and swing doors (top right), search cabinets, grasping objects and fetching it to the other robot (bottom left), operating light switches and checking lamp states (bottom right).}
    \label{fig:example-image}
    \vspace{-0.5cm}
\end{figure}
\begin{itemize}
    \item \textbf{Fetch \& Drop} This task is defined only for movable objects (e.g. bottles). As above, a grasp it estimated and Fluffy moves to the object and grasps it. The robot then drops the object into Softy's basket. To facilitate the drop, Softy lowers its base by crouching. Movements of both robots are synchronized via inter-robot communication.
    \item \textbf{Search \& Drop} This task is defined for drawers and swing doors. Fluffy first opens the drawer and uses an open vocabulary object detector~\cite{minderer2022} to check its contents. If a queried object is found, the robot proceeds to grasp it and drop it into Softy's basket. Again, Softy facilitates the drop by crouching and moving closer to Fluffy.
    \item \textbf{Operate \& Check} This task is defined for light switches within the scene. While Fluffy operates the light switch, Softy checks the states of all lamps in the scene. If Softy detects a state change in one of the lamps, a new lamp-light switch connection in the scene graph is created. The information is again relayed to the HoloLens and the scene graph gets updated with the newly found link.
\end{itemize}
\vspace{-0.1cm}
\PAR{Body Planning} In order to interact with objects without colliding with the environment or the other agent, we implement a planning algorithm that takes the other agents as a static obstacle into account. We first approximate an upper bound of the robot's spatial extend as a cuboid and render it into the 3D map. Afterwards, we generate a 2D occupancy map by projecting the 3D map of the environment onto the xy-plane (including the cuboid, excluding the floor). Given the xy-coordinates of the interactable object, we first need to generate a suitable body position that is collision free and close enough to the object to grasp it. We do this by generating a set of xy-coordinates on a circle with a $r=1$ meter radius around the object. We then select as a final position the xy-coordinates that are both closest to the current robot position in terms of euclidean distance and have a distance of at least $0.6$ meters to other obstacles. Note that more complex algorithms for planning, such as RRT* ~\cite{karaman2011sampling}, are not needed in the current setup, although they are implementable given our current map representation.

%% file: sec/4_user_study.tex
\section{User Study} %ablation study is not a good name

\begin{comment}
\begin{figure*}[h!]
    \centering
    \setlength{\tabcolsep}{1pt} % Reduce space between columns
    \renewcommand{\arraystretch}{0.8} % Reduce space between rows
    \begin{tabular}{@{}ccc@{}} % Remove padding around the table
        \includegraphics[width=0.3\textwidth]{author-kit-CVPR2025-v3.1-latex-/Images/general_eval.png} &
        \includegraphics[width=0.3\textwidth]{author-kit-CVPR2025-v3.1-latex-/Images/experience_eval.png} &
        \includegraphics[width=0.33\textwidth]{author-kit-CVPR2025-v3.1-latex-/Images/tutorial_eval.png} \\
    \end{tabular}
    \caption{\textbf{Time graphs.} Figure shows comparison between execution time for each task. From left to right we have: general times, comparison between beginner and expert times and comparison between tutorial and no tutorial execution times}
    \label{fig:time-graphs}
    \vspace{-0.5cm} % Reduce space below the figure
\end{figure*}
\end{comment}

\begin{figure*}[h!]
    \centering
    \setlength{\tabcolsep}{1pt} % Reduce space between columns
    \renewcommand{\arraystretch}{0.8} % Reduce space between rows

    % Subfigure 1
    \begin{subfigure}{0.32\textwidth}
        \centering
        \includegraphics[width=\textwidth]{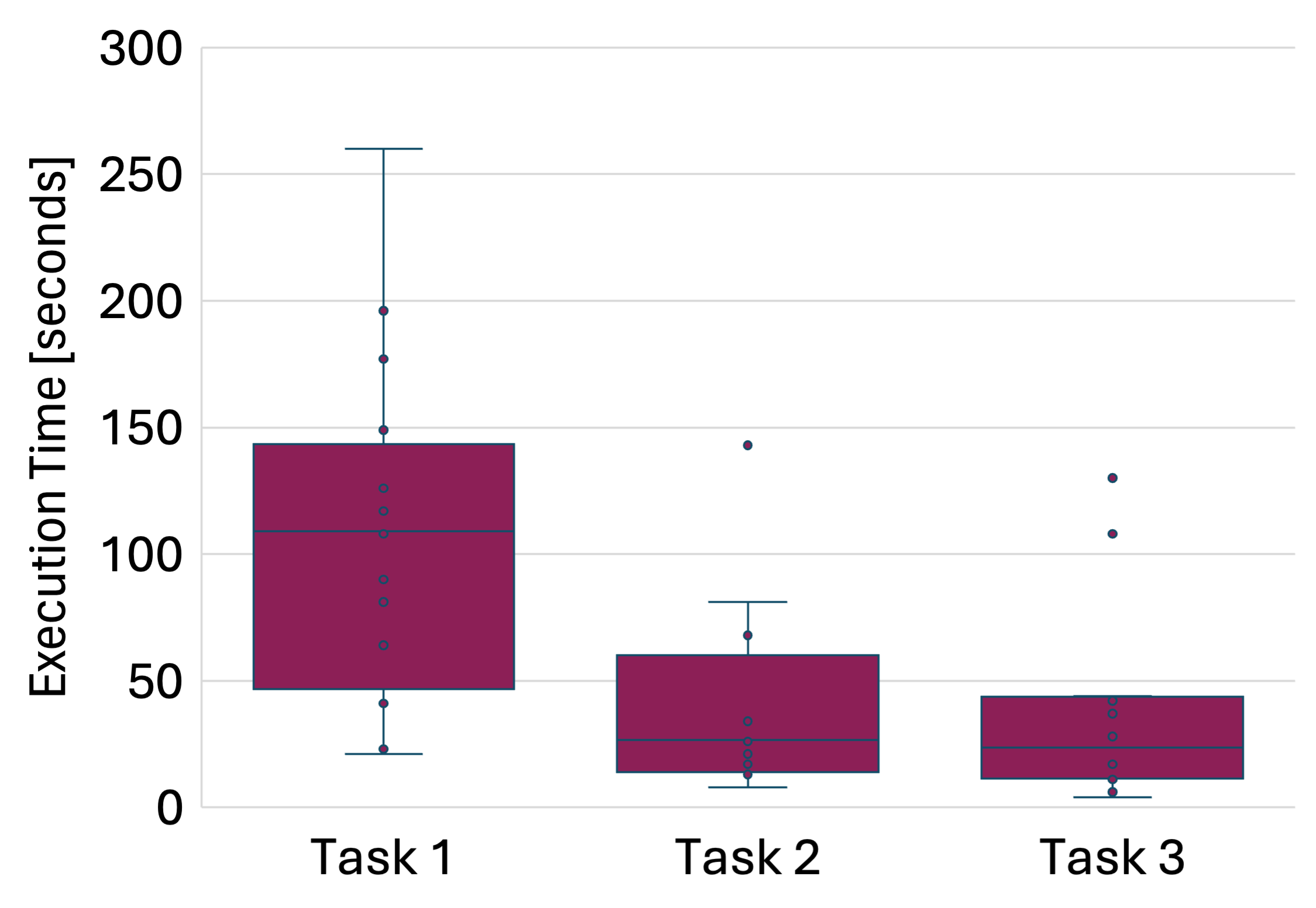}
        \caption{General times}
        \label{fig:general-times}
    \end{subfigure}
    % Subfigure 2
    \begin{subfigure}{0.32\textwidth}
        \centering
        \includegraphics[width=\textwidth]{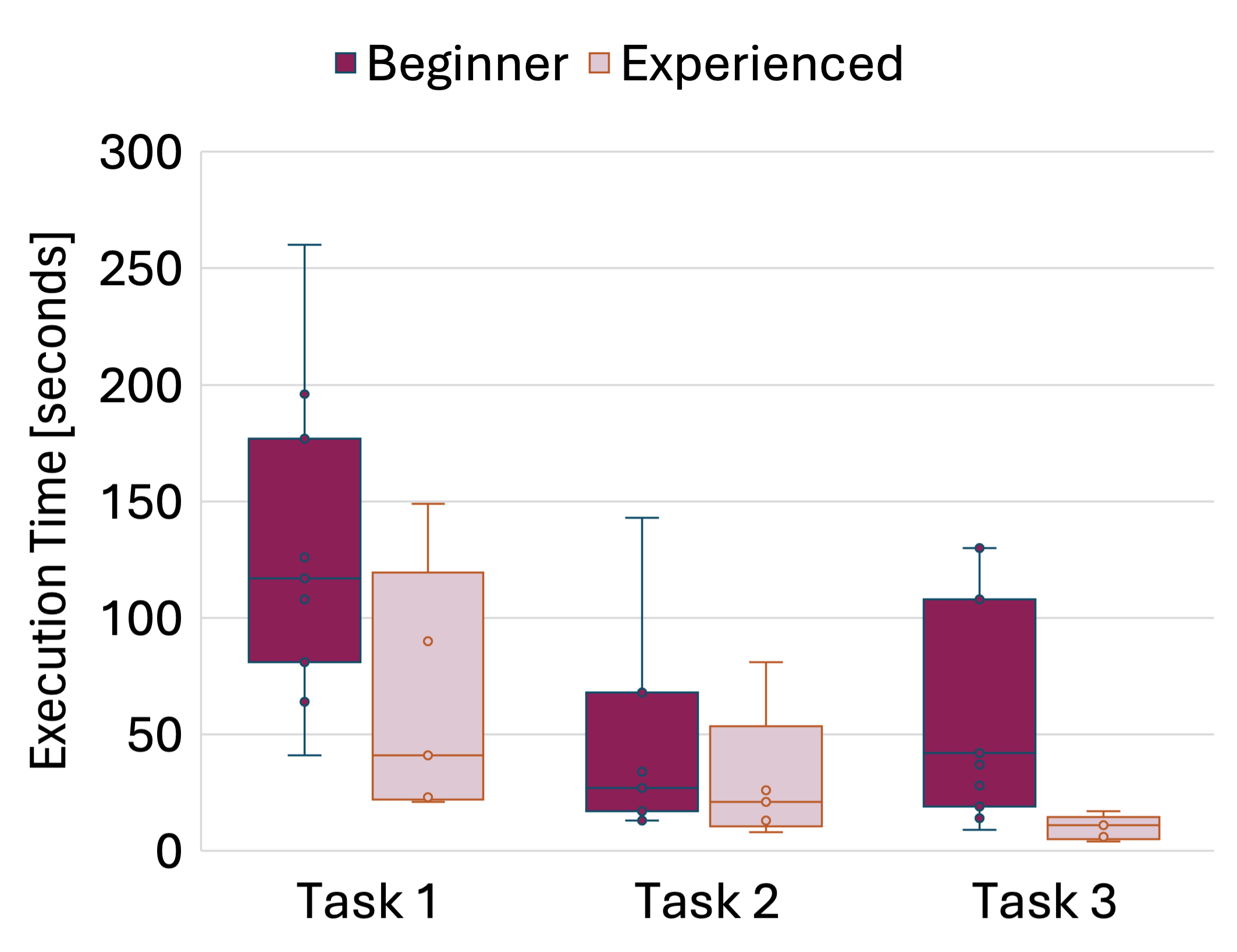}
        \caption{Beginner vs. Expert times}
        \label{fig:experience-times}
    \end{subfigure}
    % Subfigure 3
    \begin{subfigure}{0.35\textwidth}
        \centering
        \includegraphics[width=\textwidth]{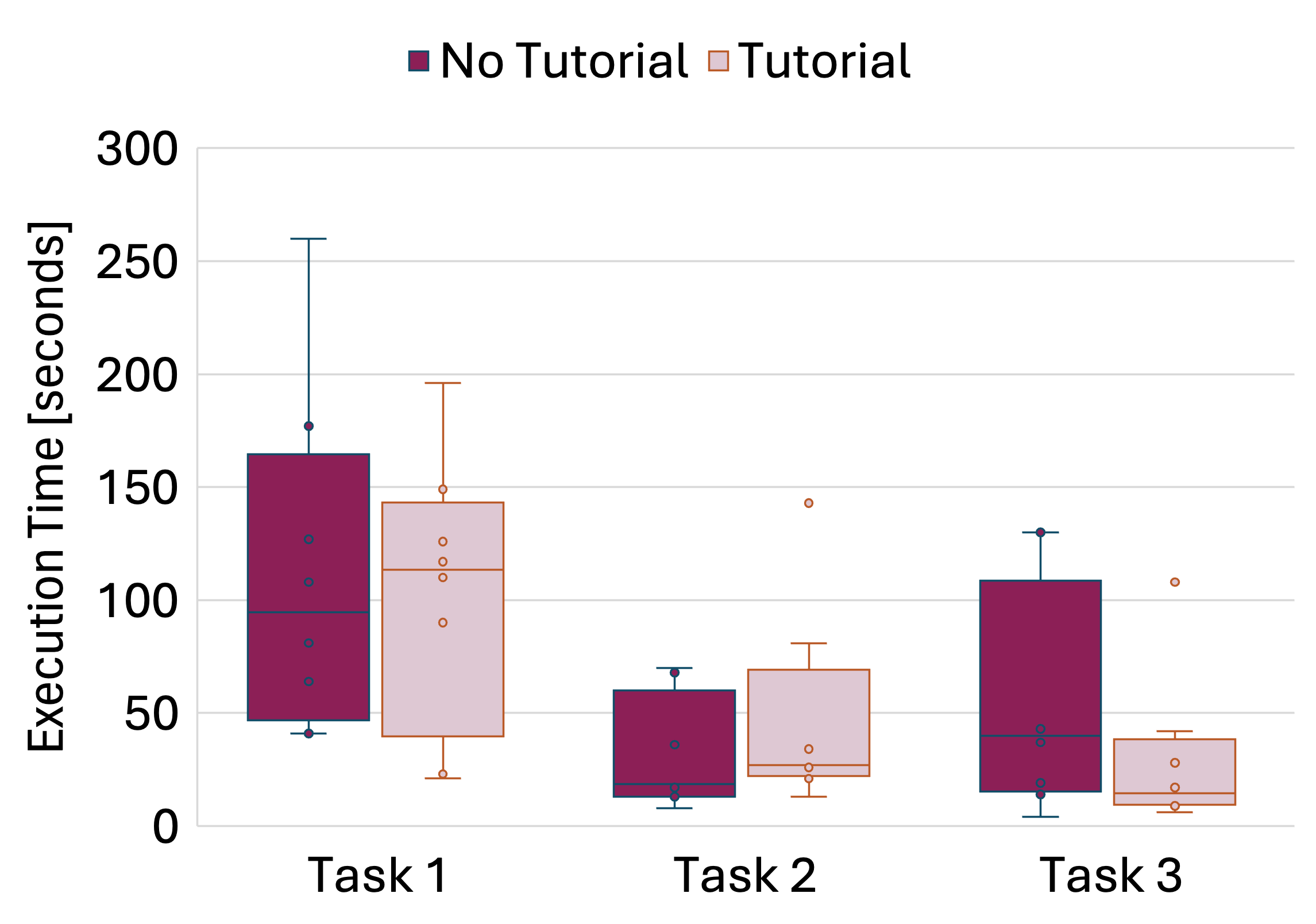}
        \caption{Tutorial vs. No Tutorial times}
        \label{fig:tutorial-times}
    \end{subfigure}

    \caption{\textbf{Time Graphs.} The figure compares execution times for three tasks. Subfigures show $25th$ to $75th$ percentile boxes with median.}
    \label{fig:time-graphs}
    \vspace{-0.3cm} % Reduce space below the figure
\end{figure*}

\vspace{-0.1cm}
%During the lifetime of the application and its development, it is of utmost importance to have real user feedback. This steers development into the desired direction as the users are the final customers of the software we create. For this reason, we use user feedback throughout the development process, leading to the final version of the application. 
Real user feedback is crucial throughout the application's development, guiding it toward the final version that meets the needs of its end users. Therefore, we separate the user study into two parts. The first part is held at our demo event. We gathered user feedback at this event to improve the our software at that time, in order to implement the final version of the application. We shall refer to this part of the user study as the pre-study. The second and primary part includes the complete user study, during which participants test our application and provide qualitative feedback on its interface and functionalities. To assess intuitiveness of the interface, we conduct quantitative tests by measuring how quickly users navigate and familiarize themselves with the application. The final study is conducted on $16$ users, with different ages ($68.8 \%$ aged $15$ to $24$, $31.2 \%$ aged $25$ to $34$) and levels of experience with MR or VR software ($68.8 \%$ without any experience with MR or VR). Our participants also come from different backgrounds, ranging from medical fields, business, computer science, food science and more. 

%This extensive study helped us polish our interface and present the final version we propose. In this section we will talk about both studies in detail and present results and feedback from our participants.
\vspace{-0.1cm}
\subsection{Pre-Study}
\label{subsec:pre-study}
\vspace{-0.1cm}
Throughout the development process, we gather informal user feedback through a pre-study to guide our functionality and design choices. The purpose of this pre-study is to increase the number of functionalities in our application and therefore give later user study participants the opportunity to test a wider range of functions. During this pre-study, users are given the opportunity to test a basic set of functions in a real scene with two robots. This is intended to give user a real life scenario example. 
\vspace{-0.1cm}
\PAR{Voice Commands} A key insight of the pre-study is that users, especially beginners, struggle with general interaction with the scene, i.e. with clicking and hovering. Therefore, to enhance the usability of the app, we implement the voice interface explained in the voice commands \cref{voice-interface}. This voice command feature reduces the need for physical interaction with the MR/VR interface during app usage. 
\vspace{-0.1cm}
\PAR{Head-Following Widgets} When designing a Mixed Reality or Virtual Reality application one has to decide on using either spatially anchored or head-following widgets (e.g. menus, message windows). While we opted for spatially anchored widgets during the earlier iterations of the design process, we now use head-following widgets in the final version of the app. This design change, driven by early user feedback, addresses the issue of users losing track of widget locations when shifting their gaze or body position. In the final version, scene, menu and prompts are in the user's focus at all times, further aiding intuitive usage of our application. This approach is further supported by our user study. That is, our participants prefer head-following widgets close to $90\%$ of the times ($88\%$).

% Our used study gives 90% for menus and prompts, when scene is considered we actualy have 40% people wanting scene to follow them. When we merged to quest scene following you was much much nicer so we stick with that. So I am nnot sure should we state this information or play dead

\begin{comment}
\begin{figure}[h!]
    \centering
   \includegraphics[width=0.7\columnwidth]{author-kit-CVPR2025-v3.1-latex-/Images/anchor_or_not.png}
    \caption{\textbf{My ass} My ass again.}
    \label{fig:anchor-or-not}
\end{figure}

I feel like adding this graphs makes not much sense since we are using a whole figure to display 4 pieces of information we already have in the text. However, humans like visuals so maybe we could leave it ...
\end{comment}

\vspace{-0.1cm}
\PAR{Further Functionalities} Lastly, the pre-study provided valuable insights and ideas for additional app functionalities desired by users. As a result, we implement a tutorial in the form of a helper window and introduce a night mode for low-light conditions. All of these newly added features will be assessed in detail in the following user study.
%Furthermore, we have introduce menus and prompts that follow user's head movement. Humans naturally move their heads a lot, which can in turn lead to losing of the menus and prompts. This means users often lose content of the app, while talking with person next to them for example. Additionally, we made the scene follow head movement in comparison to it being anchored to the start location in old app iteration. Now scene, menu and prompts are in users focus all the time, so he does not have to worry about losing it or rearranging content through the scene, which reduces number of interactions further. Both of these functionalities will be tested in our extensive user study later.
%At last, following other feedback we have implemented a tutorial (in form of helper window), high contrast night mode for low light conditions and further polished our user interface. All of the above-mentioned new functionalities we have covered in depth in previous section.
\vspace{-0.1cm}

\subsection{Main Study}
\vspace{-0.1cm}
During the main study, the users have the opportunity to test the final version of our application, including the final set of functionalities. They also get the chance to give feedback on the app's design choice and propose further improvements. We separate the main study into two parts: qualitative and quantitative analysis. This gives us feedback on design choices and intuitiveness, respectively. 
%Final study we introduce in this section gave participants the opportunity to test out our final set of features and propose further improvements and give us their feedback of design choices. 
% Dont think we need to mention that
%Study was conducted using only mixed reality headset without the robot setup. This gave us possibility to test wider audience of users considering we were not constrained to lab environment. Scene and the user interface did not differ from the realistic ones, given to users during pre-study, in any way. 
\vspace{-0.1cm}
\subsubsection{Quantitative Analysis}
\vspace{-0.1cm}
In the first part of the main study, we evaluate how intuitive the app is to users, and later investigate the utility of a tutorial as well as the impact of prior MR/VR experience. To this end, we measure the time it takes a participant to execute three tasks, which we explained to them beforehand.
%In the first part of our main study we evaluate how intuitive our app is to our users. To this end, we measure the time it takes each participant to execute 3 tasks, which we explain to them beforehand. During this section of the study we have divideusers into two groups. First group (Group A), was given a chance to have a look at the tutorial which gives them a brief understanding of the app, interactions and functionalities. Other group, or group B, did not see the tutorial and was let do dive straight into the tasks. Otherwise, tasks given to each of the groups did not differ from each other in any way. Groups were also separate equally, each group consisting of exactly $50 \%$ participant. In this section we will have a look at the results we have gathered and further discuss them.
These tasks entail navigating the user interface until a button corresponding to a free-form prompt is retrieved and operated by the participant. We select tasks of arguably the same level of difficulty. All tasks start from a common point (the start menu), and time is measured from the moment participants switch to the scene view until they press the desired command button. The tasks are listed in \cref{tab:tasks}.
\begin{table}[t]
    \centering
    \begin{tabular}{>{\raggedright\arraybackslash}p{0.2\columnwidth} p{0.7\columnwidth}}
        \toprule
        \textbf{Task} & \textbf{Description} \\ \midrule
        Task 1        & \textit{“Have the robots identify the connections between the lamps and the only light switch in the scene.”} \\ 
        Task 2        &  \textit{“Have the robots open the large white swing door of the white cabinet.”} \\
        Task 3        &  \textit{“Find the current battery level of the robot named Fluffy.”} \\ \bottomrule
    \end{tabular}
    \caption{\textbf{Tasks.} Summary of tasks performed by participants. Each participant is given the description and has to localize the button in the user interface that leads to the desired outcome.}
    \label{tab:tasks}
    \vspace{-0.7cm}
\end{table}
\vspace{-0.1cm}
\PAR{General Evaluation} Initially, we evaluate the time it takes the user to complete each task. As seen in \cref{fig:general-times}, the median completion time for task 1 is $109$ seconds (mean: $108$ seconds), while the time decreased significantly for task 2 (median: $27$ seconds, mean: $39$ seconds) and task 3 (median: $24$ seconds, mean: $41$ seconds). Assuming a comparable level of difficulty across all tasks, this indicates a steep learning curve, with the median completion time dropping considerably after just one task, highlighting the intuitiveness of our app.
\begin{comment}
 mean      median
108,19  109
38,56	  26,5
40,88	  23,5   
\end{comment}
\begin{comment}
\begin{figure}[h]
    \centering
   \includegraphics[width=1.0\columnwidth]{author-kit-CVPR2025-v3.1-latex-/Images/general_eval.png}
    \caption{\textbf{General times}. Figure shows median ($50th$ percentile), along $25th$ percentile and $75th$ percentile, time it takes each user to execute each of three tasks.}
    \label{fig:general-times}
    \vspace{-0.5cm}
\end{figure}
\end{comment}
\vspace{-0.1cm}
\PAR{Impact of Experience} We also find it insightful to compare how previous experience with MR/VR interfaces influences task completion times. As shown in \cref{fig:experience-times}, experienced users completed tasks in $64.8$, $29.8$, and $10$ seconds, with corresponding median times of $41$, $21$, and $11$ seconds. In contrast, beginner participants required an average of $127.9$, $42.6$, and $54.9$ seconds to finish each task, with median times of $117$, $27$, and $42$ seconds. One can observe that the level of experience also has a notable impact on the learning curve, with experienced users demonstrating faster learning progress. For beginner users on the other hand, the learning process appears less straightforward and progresses at a slower pace.

%It is also interesting to compare how previous experience with VR and MR interface affects times it takes to execute tasks. First of all, looking at Figures \ref{fig:general-times} and Figure \ref{fig:exp-times}, we can observe that times align for general execution times and beginner execution times. This is due to the fact that our data has less experienced users than beginner users. Experienced users took on average $64.8$, $29.8$ and $10$ seconds respectively for each task ($41$, $21$ and $11$ seconds if we consider median times). Beginner participants, on the other hand, took on average $127.9$, $42.6$ and $54.9$ seconds to complete each task ($117$, $27$ and $42$ seconds if we consider median times). Having this data in mind we can conclude that experience helps users boost execution times significantly.
\begin{comment}
mean        median

127,9090    117

64,8		41

42,5454     27	

29,8		21

54,9090	    42

10		    11    
\end{comment}

\begin{comment}
\begin{figure}[h]
    \centering
   \includegraphics[width=0.9\columnwidth]{author-kit-CVPR2025-v3.1-latex-/Images/experience_eval.png}
    \caption{\textbf{Experience comparison times}. Figure compares median ($50th$ percentile), along with $25th$ percentile and $75th$ percentile, time it takes each user to execute each of the three tasks having previous experience with the MR or VR (purple) and without experience (pink).}
    \label{fig:exp-times}
    \vspace{-0.6cm}
\end{figure}
\end{comment}

\vspace{-0.1cm}
\PAR{Impact of Tutorial} Lastly, we compare the impact of showing the tutorial on task execution times. As shown in \cref{fig:tutorial-times}, the execution times do change significantly. Specifically, participants without prior tutorial need on average $112.4$ seconds to complete task 1, compared to $104$ seconds for those participants who viewed it. For tasks 2 and 3, these times are $30$ seconds versus $46.5$ seconds, and $52.6$ seconds versus $29.1$ seconds, respectively. Participants are evenly distributed across the tutorial and non-tutorial groups. We attribute these results to two factors: the tutorial provides only textual information without interactive practice, and execution times depend more on prior experience with MR devices. As both experienced and inexperienced users participate, some perform better without the tutorial, as seen in tasks 1 and 2. Thus, the tutorial offers limited performance improvement, as execution times are primarily driven by interaction skills.
\vspace{-0.1cm}
\subsubsection{Qualitative Analysis}
\begin{figure}[h]
    \centering
    \begin{subfigure}[b]{0.49\columnwidth}
        \centering
        \includegraphics[width=\linewidth]{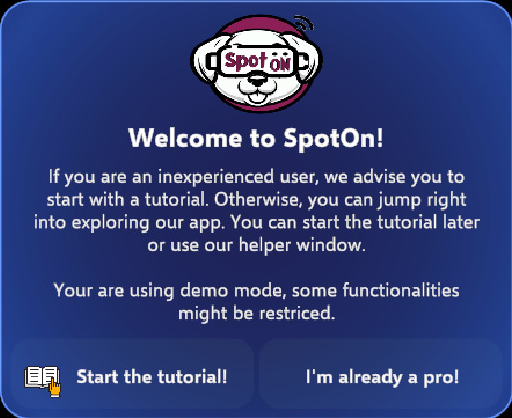}
    \end{subfigure}
    \hfill
    \begin{subfigure}[b]{0.49\columnwidth}
        \centering
        \includegraphics[width=\linewidth]{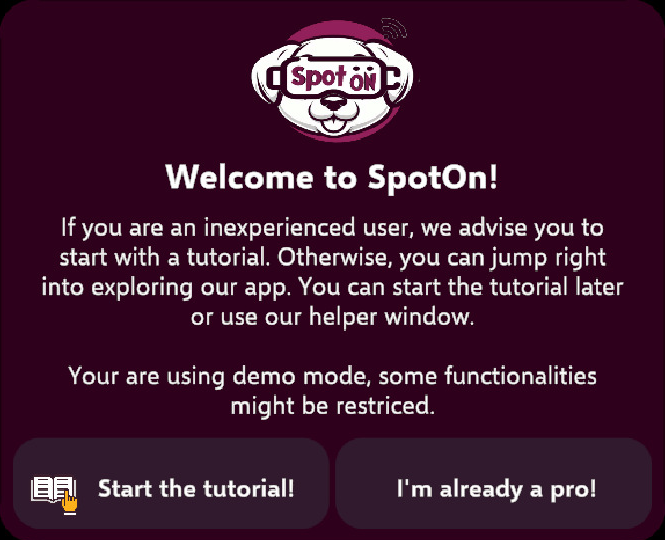}
    \end{subfigure}
    \caption{\textbf{Color scheme comparison.} Comparison of the two choices for main app color scheme.}
    \label{fig:color-scheme}
    \vspace{-0.4cm}
\end{figure}

\begin{figure}[h]
    \centering
    \begin{subfigure}[b]{0.5\columnwidth}
        \centering
        \includegraphics[width=\linewidth]{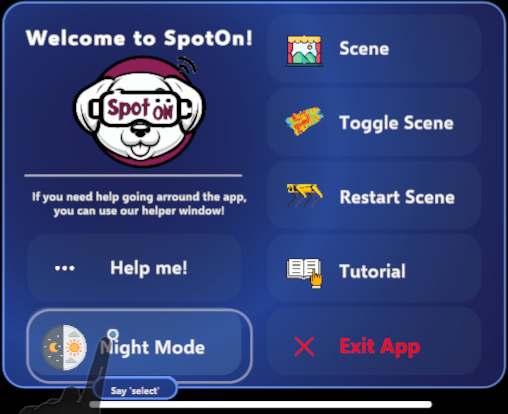}
    \end{subfigure}
    \hfill
    \begin{subfigure}[b]{0.48\columnwidth}
        \centering
        \includegraphics[width=\linewidth]{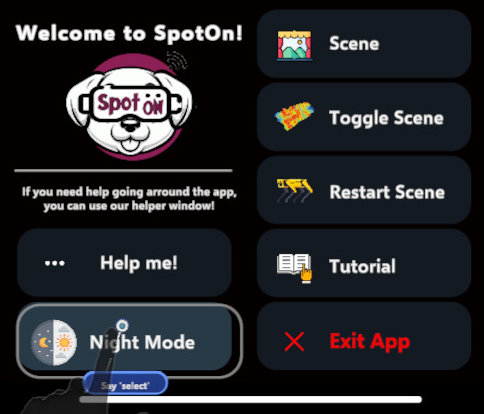}
    \end{subfigure}
    \caption{\textbf{Day/Night mode.} Comparison between night and day mode of our app.}
    \label{fig:day-night-mode}
    \vspace{-0.4cm}
\end{figure}

\begin{figure}[h]
    \centering
    \begin{subfigure}[b]{0.57\columnwidth}
        \centering
        \includegraphics[width=\linewidth]{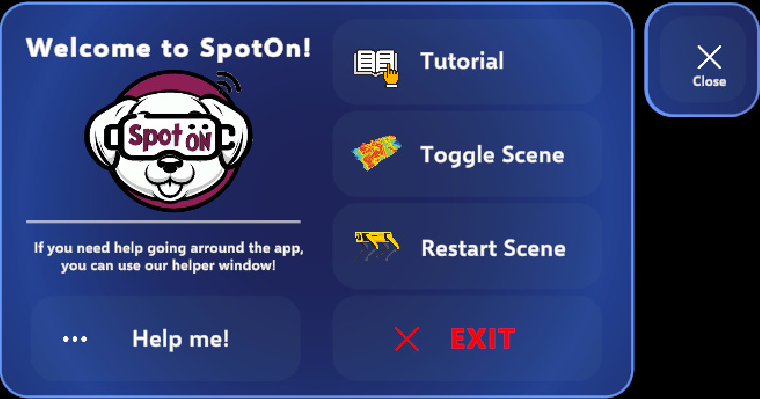}
    \end{subfigure}
    \hfill
    \begin{subfigure}[b]{0.39\columnwidth}
        \centering
        \includegraphics[width=\linewidth]{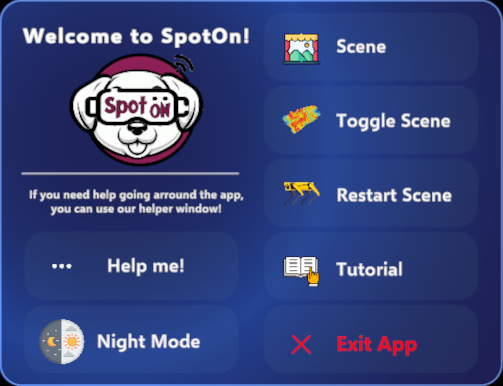}
    \end{subfigure}
    \caption{\textbf{Close and Scene button.} Comparison between two ways to transfer to scene view, close and scene button. } 
    \label{fig:close-button}
    \vspace{-0.3cm}
\end{figure}

\vspace{-0.1cm}
The second part of our user study aims at improving our user interface and design. Participants are asked for their opinion on fundamental design choice. For this, we confront the participants with pairs of different designs. We also prompt the users to rate aspects of the app such as the overall intuitiveness and design of the app. Lastly, we pose a number of open box questions, in order to give the users the freedom to name any further suggestions or complaints. 
\vspace{-0.1cm}
\PAR{Design Choices} During this part of the study participants are given three design choices shown on \cref{fig:color-scheme}, \cref{fig:day-night-mode} and \cref{fig:close-button}. Considering the color scheme of the app, the majority of users ($69 \%$) prefer the default (blue) color scheme over the purple Spot-On color scheme. A further design choice considers the transition from the main menu to the scene view. Here, the user has the choice of either using a "close" button separated from the menu or an integrated "Scene" button. A majority of $63 \%$ voted for transitioning to the scene using the integrated "Scene" button as shown in \cref{fig:close-button}. Lastly, users are asked whether they find the night mode useful, as shown in \cref{fig:day-night-mode}. An unanimous $100 \%$ of participants affirm the utility of this feature.

\vspace{-0.1cm}
\PAR{Design Grading} We prompt the users to rate different aspects of our app, such as overall visual appeal and user-friendliness. They are asked to give us feedback by giving grades on a scale of $1$ to $5$, with $1$ being "very poor" and $5$ being "very good". The exact questions and the corresponding mean scores are shown in \cref{tab:questions}. We find this feedback positive as most grades averaged above $4.2$, indicating that most participants enjoy using our interface. We would like to highlight two underperforming areas related to robot visualization (Q3) and app responsiveness (Q5). The first issue stems from the robots being static in the scene, as they were not connected to real robots during the user study. The second issue involves the app’s inconsistent response to interactions, which users report as being caused by failures in hand detection, particularly for users with smaller hands.

\begin{table}[t]
    \centering
    \small % or \footnotesize for even smaller text
    \begin{tabular}{>{\raggedright\arraybackslash}p{0.05\columnwidth} p{0.7\columnwidth} p{0.15\columnwidth}}
        \toprule
        \textbf{ID} & \textbf{Question Text} & \textbf{Mean Grade} \\ \midrule
        Q1 & How visually appealing was the interface? & $4.31$ \\
        Q2 & How intuitive was the interaction with objects and buttons in the app? & $4.25$ \\
        Q3 & Did the app respond accurately to your selections? & $3.69$ \\ 
        Q4 & How useful was the on-screen feedback provided by the app? & $4.25$ \\
        Q5 & How realistic were the robot visualizations? & $3.56$ \\
        Q6 & How would you rate the speed of the app’s responses to your inputs? & $4.69$ \\
        Q7 & How engaging was the overall experience? & $4.50$ \\ \bottomrule
    \end{tabular}
    \caption{\textbf{Questions and Grades.} This table shows text of each question presented to participants of our user study along with its corresponding grade.}
    \label{tab:questions}
    \vspace{-0.4cm}
\end{table}

\PAR{General Feedback} In the final part of our user study, participants have the opportunity to give us more open text feedback and help us improve our app with novel ideas. They find our app intuitive and well organized, with some room for improvement. This includes a less cluttered tutorial, a better clicking interface or illuminated objects when hovering over them (which we added as a feature in the final iteration of our MR/VR application). Also, because the scene visualization (which follows users head movement at all times) is less sensitive to head movement, some users perceive it as delayed. Additionally, some feedback includes the scene quality being of too low resolution or too bright. At the end, $90 \%$ of the users find our voice interface useful as it decreases number of clicks required to navigate throughout the app.

%% file: sec/5_conclusion.tex
\section{Conclusion}
\vspace{-0.1cm}
We present Spot-On, a mixed reality interface for multi-robot control, hosted on the Microsoft HoloLens 2 and Meta Quest 3. The application was developed with the help of valuable user feedback to create an intuitive and immersive experience. Key features include a digital twin of the robots' environment, interactable buttons and objects, a voice interface and a tutorial for new users. Once an action is selected in the virtual scene, two Boston Dynamics Spot robots collaboratively execute the task in the real-world environment, with live visualizations provided in the app. For future work, we propose to implement an open-vocabulary voice interface for enhanced ease-of-use, and to enable on-the-fly environment mapping using an on-board depth sensor, eliminating the need for a manually configured digital scene.
\vspace{-0.1cm}

%% file: main.bbl
\begin{thebibliography}{27}
\providecommand{\natexlab}[1]{#1}
\providecommand{\url}[1]{\texttt{#1}}
\expandafter\ifx\csname urlstyle\endcsname\relax
  \providecommand{\doi}[1]{doi: #1}\else
  \providecommand{\doi}{doi: \begingroup \urlstyle{rm}\Url}\fi

\bibitem[Behrens et~al.(2024)Behrens, Zurbrügg, Pollefeys, Bauer, and Blum]{lostfound}
Tjark Behrens, René Zurbrügg, Marc Pollefeys, Zuria Bauer, and Hermann Blum.
\newblock Lost \& found: Updating dynamic 3d scene graphs from egocentric observations, 2024.

\bibitem[Bohg et~al.(2016)Bohg, Hausman, Sankaran, Brock, Kragic, Schaal, and Sukhatme]{DBLP:journals/corr/BohgHSBKSS16}
Jeannette Bohg, Karol Hausman, Bharath Sankaran, Oliver Brock, Danica Kragic, Stefan Schaal, and Gaurav~S. Sukhatme.
\newblock Interactive perception: Leveraging action in perception and perception in action.
\newblock \emph{CoRR}, abs/1604.03670, 2016.

\bibitem[Chen et~al.(2024)Chen, Sun, Pollefeys, and Blum]{chen}
Jiaqi Chen, Boyang Sun, Marc Pollefeys, and Hermann Blum.
\newblock A 3d mixed reality interface for human-robot teaming.
\newblock In \emph{2024 IEEE International Conference on Robotics and Automation (ICRA)}, pages 11327--11333. IEEE, 2024.

\bibitem[Delitzas et~al.(2024)Delitzas, Takmaz, Tombari, Sumner, Pollefeys, and Engelmann]{scenefun3d}
Alexandros Delitzas, Ayca Takmaz, Federico Tombari, Robert Sumner, Marc Pollefeys, and Francis Engelmann.
\newblock {SceneFun3D: Fine-Grained Functionality and Affordance Understanding in 3D Scenes}.
\newblock In \emph{CVPR}, 2024.

\bibitem[Engelbracht et~al.(2024)Engelbracht, Zurbrügg, Pollefeys, Blum, and Bauer]{engelbracht2024}
Tim Engelbracht, René Zurbrügg, Marc Pollefeys, Hermann Blum, and Zuria Bauer.
\newblock Spotlight: Robotic scene understanding through interaction and affordance detection, 2024.

\bibitem[Fang et~al.(2023)Fang, Wang, Fang, Gou, Liu, Yan, Liu, Xie, and Lu]{fang2023}
Hao-Shu Fang, Chenxi Wang, Hongjie Fang, Minghao Gou, Jirong Liu, Hengxu Yan, Wenhai Liu, Yichen Xie, and Cewu Lu.
\newblock Anygrasp: Robust and efficient grasp perception in spatial and temporal domains, 2023.

\bibitem[Garcia et~al.(2024)Garcia, Lukovic, Reynaud, Sgobbi, Bruni, Brun, Z{\"u}nd, Bollati, Pollefeys, Blum, et~al.]{holospot}
Pablo~Soler Garcia, Petar Lukovic, Lucie Reynaud, Andrea Sgobbi, Federica Bruni, Martin Brun, Marc Z{\"u}nd, Riccardo Bollati, Marc Pollefeys, Hermann Blum, et~al.
\newblock Holospot: Intuitive object manipulation via mixed reality drag-and-drop.
\newblock \emph{arXiv preprint arXiv:2410.11110}, 2024.

\bibitem[Gu et~al.(2023)Gu, Kuwajerwala, Morin, Jatavallabhula, Sen, Agarwal, Rivera, Paul, Ellis, Chellappa, Gan, {de Melo}, Tenenbaum, Torralba, Shkurti, and Paull]{conceptgraphs}
Qiao Gu, Alihusein Kuwajerwala, Sacha Morin, {Krishna Murthy} Jatavallabhula, Bipasha Sen, Aditya Agarwal, Corban Rivera, William Paul, Kirsty Ellis, Rama Chellappa, Chuang Gan, {Celso Miguel} {de Melo}, {Joshua B.} Tenenbaum, Antonio Torralba, Florian Shkurti, and Liam Paull.
\newblock Conceptgraphs: Open-vocabulary 3d scene graphs for perception and planning.
\newblock \emph{arXiv}, 2023.

\bibitem[Hausman et~al.(2015)Hausman, Niekum, Osentoski, and Sukhatme]{7139655}
Karol Hausman, Scott Niekum, Sarah Osentoski, and Gaurav~S. Sukhatme.
\newblock Active articulation model estimation through interactive perception.
\newblock In \emph{2015 IEEE International Conference on Robotics and Automation (ICRA)}, pages 3305--3312, 2015.

\bibitem[Hsu et~al.(2023)Hsu, Jiang, and Zhu]{hsu2023dittohousebuildingarticulation}
Cheng-Chun Hsu, Zhenyu Jiang, and Yuke Zhu.
\newblock Ditto in the house: Building articulation models of indoor scenes through interactive perception, 2023.

\bibitem[Iglesius et~al.(2024)Iglesius, Kobayashi, Uranishi, and Takemura]{iglesius2024mrnabmixedrealitybasedrobot}
Eduardo Iglesius, Masato Kobayashi, Yuki Uranishi, and Haruo Takemura.
\newblock Mrnab: Mixed reality-based robot navigation interface using optical-see-through mr-beacon, 2024.

\bibitem[Jiang et~al.(2022)Jiang, Hsu, and Zhu]{jiang2022dittobuildingdigitaltwins}
Zhenyu Jiang, Cheng-Chun Hsu, and Yuke Zhu.
\newblock Ditto: Building digital twins of articulated objects from interaction, 2022.

\bibitem[Jocher et~al.(2023)Jocher, Qiu, and Chaurasia]{Jocher_Ultralytics_YOLO_2023}
Glenn Jocher, Jing Qiu, and Ayush Chaurasia.
\newblock {Ultralytics YOLO}, 2023.

\bibitem[Karaman and Frazzoli(2011)]{karaman2011sampling}
Sertac Karaman and Emilio Frazzoli.
\newblock Sampling-based algorithms for optimal motion planning.
\newblock \emph{The International Journal of Robotics Research}, 30\penalty0 (7):\penalty0 846--894, 2011.

\bibitem[Lemke et~al.(2024)Lemke, Bauer, Zurbrügg, Pollefeys, Engelmann, and Blum]{lemke2024}
Oliver Lemke, Zuria Bauer, René Zurbrügg, Marc Pollefeys, Francis Engelmann, and Hermann Blum.
\newblock Spot-compose: A framework for open-vocabulary object retrieval and drawer manipulation in point clouds, 2024.

\bibitem[Liu et~al.(2024)Liu, Orru, Vakil, Paxton, Shafiullah, and Pinto]{ok-robot}
Peiqi Liu, Yaswanth Orru, Jay Vakil, Chris Paxton, Nur Shafiullah, and Lerrel Pinto.
\newblock Demonstrating ok-robot: What really matters in integrating open-knowledge models for robotics.
\newblock In \emph{Robotics: Science and Systems XX}. Robotics: Science and Systems Foundation, 2024.

\bibitem[Martin-Martin and Brock(2014)]{inproceedings}
Roberto Martin-Martin and Oliver Brock.
\newblock Online interactive perception of articulated objects with multi-level recursive estimation based on task-specific priors.
\newblock 2014.

\bibitem[Microsoft()]{hololens}
Microsoft.
\newblock Hololens.
\newblock \url{https://www.microsoft.com/en-us/hololens}.
\newblock Accessed: 2025-01-04.

\bibitem[Minderer et~al.(2022)Minderer, Gritsenko, Stone, Neumann, Weissenborn, Dosovitskiy, Mahendran, Arnab, Dehghani, Shen, Wang, Zhai, Kipf, and Houlsby]{minderer2022}
Matthias Minderer, Alexey Gritsenko, Austin Stone, Maxim Neumann, Dirk Weissenborn, Alexey Dosovitskiy, Aravindh Mahendran, Anurag Arnab, Mostafa Dehghani, Zhuoran Shen, Xiao Wang, Xiaohua Zhai, Thomas Kipf, and Neil Houlsby.
\newblock Simple open-vocabulary object detection with vision transformers, 2022.

\bibitem[Nie et~al.(2023)Nie, Gadre, Ehsani, and Song]{nie2023structureactionlearninginteractions}
Neil Nie, Samir~Yitzhak Gadre, Kiana Ehsani, and Shuran Song.
\newblock Structure from action: Learning interactions for articulated object 3d structure discovery, 2023.

\bibitem[Nunes and Demiris(2022)]{10.1109/ICRA46639.2022.9812430}
Urbano~Miguel Nunes and Yiannis Demiris.
\newblock Kinematic structure estimation of arbitrary articulated rigid objects for event cameras.
\newblock In \emph{2022 International Conference on Robotics and Automation (ICRA)}, page 508–514. IEEE Press, 2022.

\bibitem[OpenAI(2025)]{OpenAI2025}
OpenAI.
\newblock Chatgpt api, 2025.

\bibitem[Radford et~al.(2021)Radford, Kim, Hallacy, Ramesh, Goh, Agarwal, Sastry, Askell, Mishkin, Clark, Krueger, and Sutskever]{clip}
Alec Radford, Jong~Wook Kim, Chris Hallacy, Aditya Ramesh, Gabriel Goh, Sandhini Agarwal, Girish Sastry, Amanda Askell, Pamela Mishkin, Jack Clark, Gretchen Krueger, and Ilya Sutskever.
\newblock Learning transferable visual models from natural language supervision, 2021.

\bibitem[Rosinol et~al.(2020)Rosinol, Gupta, Abate, Shi, and Carlone]{rosinol20203ddynamicscenegraphs}
Antoni Rosinol, Arjun Gupta, Marcus Abate, Jingnan Shi, and Luca Carlone.
\newblock 3d dynamic scene graphs: Actionable spatial perception with places, objects, and humans, 2020.

\bibitem[Schult et~al.(2023)Schult, Engelmann, Hermans, Litany, Tang, and Leibe]{mask3d}
Jonas Schult, Francis Engelmann, Alexander Hermans, Or Litany, Siyu Tang, and Bastian Leibe.
\newblock {Mask3D: Mask Transformer for 3D Semantic Instance Segmentation}.
\newblock 2023.

\bibitem[Takmaz et~al.(2023)Takmaz, Fedele, Sumner, Pollefeys, Tombari, and Engelmann]{openmask3d}
Ay{\c{c}}a Takmaz, Elisabetta Fedele, Robert~W Sumner, Marc Pollefeys, Federico Tombari, and Francis Engelmann.
\newblock Openmask3d: Open-vocabulary 3d instance segmentation.
\newblock \emph{arXiv preprint arXiv:2306.13631}, 2023.

\bibitem[Wang et~al.(2024)Wang, Chen, Yu, Xu, Chen, Fu, Lu, Mu, and Luo]{wang2024articulated}
Xi Wang, Tianxing Chen, Qiaojun Yu, Tianling Xu, Zanxin Chen, Yiting Fu, Cewu Lu, Yao Mu, and Ping Luo.
\newblock Articulated object manipulation using online axis estimation with sam2-based tracking.
\newblock \emph{arXiv preprint arXiv:2409.16287}, 2024.

\end{thebibliography}
